\documentclass[pra,twocolumn,aps,amssymb,showpacs,superscriptaddress]{revtex4-1}
\usepackage{epsfig}
\usepackage{graphicx}
\usepackage{amsmath}
\usepackage{amssymb}
\usepackage{amsmath,amssymb}
\usepackage{enumerate}
\usepackage{hyperref}
\usepackage[normalem]{ulem}
\usepackage{cancel}
\usepackage{comment}
\usepackage{color}
\definecolor{rosso}{rgb}{1,0,0}
\definecolor{verde}{rgb}{0,1,0}
\definecolor{blue}{rgb}{0,0,1}
\definecolor{verdescuro}{rgb}{0,0.5,0.5}
\definecolor{rossoscuro}{rgb}{0.7,0.3,0}
\definecolor{bluscuro}{rgb}{0.3,0,0.7}
\definecolor{magenta}{rgb}{1,0,1}
\newcommand{\be}{\begin{equation}}
\newcommand{\bea}{\begin{eqnarray}}
\newcommand{\eea}{\end{eqnarray}}
\newcommand{\ee}{\end{equation}}
\begin{document}

\title{Evolution of an attractive polarized Fermi gas: \\ From a Fermi liquid of polarons to a non-Fermi liquid \\at the Fulde-Ferrell-Larkin-Ovchinnikov quantum critical point}

\author{M. Pini}
\email{pini@pks.mpg.de}
\affiliation{Max Planck Institute for the Physics of Complex Systems, N\"{o}thnitzer Str. 38, 01187 Dresden, Germany}
\affiliation{CNR-INO, Istituto Nazionale di Ottica, Sede di Firenze, 50125 (FI), Italy}
\affiliation{School of Science and Technology, Physics Division, Universit\`{a} di Camerino, 62032 Camerino (MC), Italy}
\author{P. Pieri}
\email{pierbiagio.pieri@unibo.it}
\affiliation{Dipartimento di Fisica e Astronomia, Universit\`{a} di Bologna, I-40127 Bologna (BO), Italy}
\affiliation{INFN, Sezione di Bologna, I-40127 Bologna (BO), Italy}
\author{G. Calvanese Strinati}
\email{giancarlo.strinati@unicam.it}
\affiliation{School of Science and Technology, Physics Division, Universit\`{a} di Camerino, 62032 Camerino (MC), Italy}
\affiliation{CNR-INO, Istituto Nazionale di Ottica, Sede di Firenze, 50125 (FI), Italy}

\begin{abstract}
The evolution of an attractive polarized two-component Fermi gas at zero temperature is analyzed as its polarization is progressively decreased, from full polarization (corresponding to the polaronic limit) 
down to a critical polarization when superfluidity sets in.  
This critical polarization and the nature of the associated superfluid instability are determined within a fully self-consistent $t$-matrix approach implemented \emph{exactly\/} at zero temperature.
In this way, the polarization-vs-coupling phase diagram at zero temperature is constructed throughout the whole BCS-BEC crossover.
Depending on the coupling strength of the inter-particle interaction between the two components, the superfluid instability can be either toward a Fulde-Ferrel-Larkin-Ovchinnikov (FFLO) phase 
or toward a standard polarized BCS phase.  
The evolution with polarization of the quasi-particle parameters in the normal Fermi gas turns out to be notably different in the two cases.
When the instability is toward a polarized BCS superfluid, quasi-particles in the proximity of the two Fermi surfaces remain well defined for all polarizations. 
When the instability is instead toward an FFLO superfluid, \emph{precursor effects\/} become apparent upon approaching the FFLO quantum critical point (QCP), where the quasi-particle residues vanish and the effective masses diverge.
This behavior leads to a complete breakdown of the quasi-particle picture characteristic of a Fermi liquid, similarly to what occurs in heavy-fermion materials at an antiferromagnetic QCP.
At unitarity, the system is further investigated at finite temperature, making it possible to identify a non-Fermi liquid region in the temperature-vs-polarization phase diagram associated with the underlying FFLO QCP. 

\end{abstract}

\maketitle


\section{Introduction}
\label{sec:introduction}
The Fermi liquid theory, as originally developed by Landau for describing liquid $^3$He \cite{Landau-1956,Landau-1957,Nozieres-1964}, is one of the most successful theories in condensed-matter physics, wherein 
it describes the behavior of metals or compounds in terms of low-energy excitations of weakly interacting fermionic quasi-particles. 
Notable exceptions to the Fermi liquid description have, however, emerged in nature, for instance, in underdoped cuprates \cite{Lee-2006} or heavy-fermion materials \cite{Stewart-2001}. 
Non-Fermi liquid behaviors are often driven by the proximity to a quantum critical point (QCP) and considerable theoretical efforts have been made to describe them \cite{Varma-2002,Lohneysen-2007,Senthil-2008}, 
also in the light of possible connections with high-temperature superconductivity \cite{Lee-2006}.

In this context, ultra-cold Fermi gases offer new perspectives for investigating non-Fermi liquid behaviors close to a QCP, owing to the large degree of control achieved in their experimental realizations.
These physical systems can be assimilated to a spin-$1/2$ Fermi gas with an attractive contact interaction. 
In this system, when a perfect matching occurs between the populations of the two spin components, fermion pairs are formed that condense to a homogeneous superfluid phase at sufficiently low temperature. 
By varying the strength of the attractive interaction via a Fano-Feshbach resonance \cite{Chin-2010}, this superfluid phase crosses over from a Bardeen-Cooper-Schrieffer (BCS) condensate of highly overlapping Cooper pairs 
in weak coupling, to a Bose-Einstein condensate (BEC) of dilute tightly-bound dimers in strong coupling, passing through an intermediate (unitary) regime of interaction associated with a divergent s-wave scattering length 
(cf. Refs.~\cite{Zwerger-2012,Physics-Reports-2018} for a review). 
 
Quite generally, the occurrence of an imbalance between the two spin populations (i.e., of a finite polarization) hinders pairing and thus superfluidity. 
For sufficiently large polarization and not too strong an attraction, the system remains in the normal phase even at zero temperature.
This phase is expected to be well described by the Fermi liquid theory \cite{Lobo-2006b,Nascimbene-2011}, even though the inter-particle interaction is now attractive and not repulsive as it was assumed in the original
formulation of the theory \cite{Landau-1956,Landau-1957,Nozieres-1964}. 
In this phase, there are two kinds of quasi-particles that correspond to fermions belonging, respectively, to the majority or minority spin species dressed by the interaction with fermions of the other species. 
In the ultra-cold gases community, when the polarization is large enough the few dressed minority fermions that are around are regarded as mobile ``impurities" embedded in the Fermi sea of the other component, 
and are usually referred to as ``polarons''.
Accordingly, in this limit the Fermi liquid itself is regarded as a dilute gas of this kind of polarons (cf. Ref.~\cite{Massignan-2014} for a review).

In the following, we shall determine how this real-space picture evolves with continuity to a more conventional Fermi liquid description in terms of quasi-particle arising in momentum space around the Fermi surface(s).
This evolution will further be extended to the point where \emph{precursor effects\/} will signal the collapse of the Fermi liquid itself already in the normal phase.
In this way, we shall be able to provide a unifying description between the approaches to a Fermi liquid adopted in the ultra-cold gases and condensed-matter communities.

Specifically, by decreasing the spin polarization from the polaronic limit, the properties of the normal Fermi liquid are expected to vary continuously down to a critical polarization, 
where a phase transition toward a superfluid phase occurs and the Fermi liquid description ceases to be valid. 
Quite generally, the superfluid phase transition can be either of first order or continuous. 
For a first-order transition, the transition point obtained by assuming a homogeneous system will be surmounted by a phase separation between a balanced superfluid and a polarized normal phase. 
In this case, the QCP will be masked by the phase separation region, although its effects could still extend beyond this region. 
For a continuous transition, the QCP will instead be directly accessible, such that possible deviations from Fermi liquid theory could more readily be observed in the proximity of the QCP. 
In addition, the continuous transition could be either toward a polarized BCS superfluid (known as Sarma state from the original work by Sarma \cite{Sarma-1963}), or toward a Fulde-Ferrel-Larkin-Ovchinnikov 
(FFLO) superfluid \cite{FF-1964,LO-1964} whereby pairs condense with a finite value of center-of-mass momentum to compensate for the mismatch of the Fermi surfaces.
In fact, it will be shown below (cf. in particular Fig.~\ref{Figure-2}(b)) that the occurrence of either one of these zero-temperature phase transitions depends on the strength of the inter-particle interaction, 
that we shall allow to span across the BCS-BEC crossover.

Particularly in the context of the FFLO superfluid phase, a close connection can naturally be established with the phenomenology occurring in condensed matter, where it is the Zeeman splitting due to a magnetic field acting on the electron spins to produce the spin imbalance that may possibly give rise to an FFLO phase.
Accordingly, the precursor effects, that we shall determine below to occur in the normal phase above the FFLO phase with regards to ultra-cold Fermi gases, could also provide hints about the long sought search for an FFLO phase in condensed matter. In this way, the critical polarization that we will obtain here over a wide spectrum of the system parameters will correspond to the ``upper'' critical field in condensed matter at which the FFLO phase breaks down.

As a matter of fact, the behavior of the polarized normal Fermi gas when approaching an FFLO QCP is of particular interest. By using renormalization group (RG) arguments, it was pointed out in Ref.~\cite{James-2010} that in dimensions $D < 3$ the quasi-particle residues vanish at the FFLO QCP, similarly to what happens at an antiferromagnetic QCP in heavy-fermion materials \cite{Varma-2002,Lohneysen-2007,Senthil-2008}. 
Deviations from Fermi liquid theory close to the FFLO QCP in anisotropic 2D systems were also found in Refs.~\cite{Piazza-2016,Pimenov-2018} within a paring fluctuations and an RG approach, respectively, 
while Ref.~\cite{Samokhin-2006} predicted a non-Fermi-liquid quasi-particle lifetime at the FFLO QCP both in 3D and 2D within a pairing fluctuations approach.

All these previous works \cite{James-2010,Piazza-2016,Pimenov-2018,Samokhin-2006} were formulated with the following restrictions: 
(i) The attractive inter-particle interaction was restricted to the weak-coupling limit; 
(ii) The theory was formulated only in the proximity of the QCP; 
(iii) The explicit dependence of physical properties (like the position of the QCP, its character, and the extension of the critical region) on the parameters of a microscopic Hamiltonian was not addressed. 
 All these restrictions will altogether be avoided in the present work, thereby allowing us to determine in a complete and consistent way the evolution of an attractive polarized Fermi gas
from the polaronic limit to the superfluid QCP.

Specifically, we will consider a microscopic Hamiltonian describing an ultra-cold two-component Fermi gas in the presence of a broad Fano-Feshbach resonance \cite{Simonucci-2005}, for which the effective interaction can be parametrized 
only in terms of the scattering length $a_{\rm F}$ between two fermions in vacuum. 
For this system, we will first obtain the critical polarization \emph{at zero temperature\/} as a function of coupling for the continuous transition from the normal to the superfluid phase, and determine explicitly whether the underlying superfluid phase is an FFLO or a polarized BCS phase. 
We will then follow the evolution of the Fermi liquid normal phase from the polaronic limit of maximum polarization down to the QCP where the Fermi liquid theory breaks down. 
For the unitary Fermi gas, we will also study the system at finite temperature and determine the extension of the non-Fermi-liquid region about the QCP. 

Our numerical calculations rely on a fully self-consistent $t$-matrix (also known as Luttinger-Ward) approach~\cite{Haussmann-1993,Haussmann-1994,Haussmann-2007,PPS-2019}.  
This approach compares well with experimental data and Quantum Monte Carlo (QMC) calculations for several thermodynamic quantities at unitarity in the balanced case~\cite{Sommer-2012,Zwerger-2016,Carcy-2019,Mukherjee-2019,Jensen-2020,Rammelmueller-2021}, as well as with QMC calculations in the polarized case~\cite{Goulko-2010,Rammelmueller-2020}. 
In this context, a recent work by two of us \cite{Pieri-2017} has formally proven that the self-consistent $t$-matrix approach exactly satisfies the Luttinger theorem~\cite{Luttinger-1960} for each Fermi surface of the two spin components. 
This property turns out to be particularly important for describing the spin-imbalanced Fermi liquid phase at zero temperature in a consistent way.
 
It should be emphasized that our zero-temperature results are obtained by implementing numerically the self-consistent $t$-matrix approach \emph{exactly\/} at $T=0$, 
thereby avoiding a $T \to 0$ extrapolation like that adopted by more conventional finite-temperature calculations.
This novel implementation constitutes a non-trivial task, which is anyhow required to correctly identify key features occurring in Fermi-liquid theory and related to the sharpness of the underlying Fermi surface,
that would otherwise be blurred away by finite-temperature effects.
In addition, considering exactly $T=0$ is required to correctly identify the FFLO critical line, which by our approach is found to exist only at $T=0$ \cite{Pini-PRR-2021} due to the inherent instability of FFLO long-range order to thermal fluctuations \cite{Shimahara-1998,Shimahara-1999,Ohashi-2002,Radzihovsky-2011,Jakubczyk-2017,Wang-2018,Zdybel-2021}.

Our previous work, that focused on the pairing susceptibility \cite{Pini-PRR-2021}, has shown that, in the unitary regime of interaction, precursor effects of the FFLO pairing fluctuations are present at finite temperature over an extended range of polarizations (as also recently pointed out in Ref.~\cite{Diessel-2022} at zero temperature). Here, we instead investigate the effects induced by these FFLO fluctuations on Fermi liquid properties and their ultimate breakdown in the proximity of an FFLO QCP.

The main results of the present article can be summarized as follows: 
(i) We have determined the polarization-vs-coupling phase diagram at zero temperature where, by comparing with available QMC and experimental data on phase separation, we find that a FFLO phase 
should indeed occur at zero temperature from weak coupling to just past unitarity. 
(ii) We have characterized the evolution of the quasi-particle residues and effective masses for the two spin components as functions of the spin-polarization at zero temperature for various couplings. 
This highlights the difference between a QCP toward a polarized BCS superfluid around which quasi-particles remain well defined, and a QCP toward an FFLO superfluid around which a quasi-particle 
description breaks down. 
(iii) We have analyzed the self-energies at zero temperature with emphasis on deviations from Fermi liquid theory close to the FFLO QCP, which is found to be consistent with a dynamical critical exponent $z=2$; 
(iv) We have identified a non-Fermi liquid critical region above the FFLO QCP in the temperature-vs-polarization phase diagram at unitarity, which is delimited by a crossover temperature $T^*_\text{NFL}$ 
determined by comparing the time-scales of thermal and quantum fluctuations close to the QCP.

The article is organized as follows. 
Section~\ref{sec:theo} describes the microscopic model and the main equations of the self-consistent $t$-matrix approach at zero temperature. 
Section~\ref{sec:Luttinger} presents, as a first check on our numerical calculations, results for the momentum distributions which show an explicit consistency with the Luttinger theorem, 
in contrast with the results obtained within a non-self-consistent $t$-matrix approach. 
Section~\ref{sec:phase_diagram} discusses the polarization-vs-coupling phase diagram at zero temperature and compares it with available experimental and QMC data.
Section~\ref{sec:Evolution_FL} characterizes the evolution with polarization of the Fermi liquid phase at zero temperature, by presenting numerical results for the quasi-particle residues and effective masses of the two spin components, and discusses the scaling of dynamical quantities close to the FFLO QCP.
Section~\ref{sec:finite_temperature} analyses the non-Fermi liquid effects at finite temperature due to the proximity to the FFLO QCP, focusing on the unitary Fermi gas.  
Section~\ref{sec:conclusions} gives our conclusions. 
In addition, Appendix~\ref{app:details_numerical_T=0} provides details of the zero-temperature algorithm for the self-consistent $t$-matrix approach, and 
Appendix~\ref{app:failure_NSC} discusses a shortcoming of the non-self-consistent $t$-matrix approach in the spin-imbalanced case at zero temperature, which is overcome by the self-consistent approach. 


\section{Theoretical formalism}
\label{sec:theo}
We consider a system of spin-$1/2$ fermions of mass $m$ mutually interacting through an attractive contact interaction, as described by the Hamiltonian (in the following, the reduced Planck constant $\hbar$ and the Boltzmann constant $k_B$ are set equal to unity): 
 \bea
\hat{H}&=&\sum_{\sigma}\int \! d {\bf r} \, \hat{\psi}^{\dagger}_{\sigma}({\bf r}) \left( - \frac{\nabla^2}{2m} \right) \hat{\psi}_{\sigma}({\bf r}) \nonumber \\
&+& v_0\int \! d {\bf r} \, \hat{\psi}^{\dagger}_{\uparrow}({\bf r}) \hat{\psi}^{\dagger}_{\downarrow}({\bf r}) \hat{\psi}_{\downarrow}({\bf r}) \hat{\psi}_{\uparrow}({\bf r}) \, .
\eea
 Here, $\hat{\psi}_{\sigma}({\bf r})$ is a field operator with spin projection \mbox{$\sigma=(\uparrow,\downarrow)$} and $v_0 < 0$ is the bare interaction strength 
 (with $v_0\to0^{-}$ when the contact interaction is regularized in terms of the two-fermion scattering length $a_{\rm F}$~\cite{Pieri-2000}).
 
Quite generally, the single-particle Green's function $G_\sigma(k)$ for  spin $\sigma$ can be expressed in terms of the self-energy $\Sigma_{\sigma}(k)$ through the Dyson equation
\begin{equation}
G_\sigma(k) = \Big[G_{0\sigma}(k)^{-1} -\Sigma_{\sigma}(k)\Big]^{-1},
\label{eq:G_k}
\end{equation}
where $G_{0\sigma}(k)=[i \omega -\mathbf{k}^2/(2m)+\mu_\sigma]^{-1}$ is the non-interacting counterpart, $m$ the fermion mass, and $\mu_\sigma$ the chemical potential of the $\sigma$ component, 
and we are adopting the four-vector notation $k=({\bf k},i\omega)$.  

We emphasize that, in the present work, the theory is formulated \emph{strictly at zero temperature\/}, upon taking the $T\to 0$ limit beforehand in all relevant equations obtained within the finite-temperature 
(Matsubara) formalism. 
Accordingly, discrete fermionic Matsubara frequencies $i \omega_n=i (2 n+1)\pi T$  ($n$ integer) or bosonic Matsubara frequencies $i \Omega_\nu = i 2 \nu \pi T$  ($\nu$ integer) are replaced by continuous 
frequencies $i \omega$ and $i \Omega$ along the imaginary axis, and the discrete sums $T \, \Sigma_n$ or $T \, \Sigma_\nu$ are replaced by integrals $\int d \omega/(2\pi)$ and $\int d \Omega/(2\pi)$.
The advantage of working with imaginary frequencies even at $T=0$ is that, in this way, one avoids the singularities (or near singularities) occurring in the Green's functions when calculated along the real frequency axis. 

The self-consistent $t$-matrix approach is then implemented at zero temperature by taking for the self-energy the expression
\begin{equation}
\Sigma_\sigma(k) = - \int \!\! \frac{d{\bf Q}}{(2\pi)^3} \int \!\! \frac{d{\Omega}}{2\pi} \, \Gamma(Q) \, G_{\bar{\sigma}}(Q-k) 
\label{eq:Sigma_k}
\end{equation}
with $Q=({\mathbf{Q},i \Omega})$, where the Green's function for fermions with opposite spin $\bar{\sigma}$ is convoluted with the particle-particle (pair) propagator
\begin{equation}
\Gamma(Q) = - \bigg[\frac{m}{4\pi a_{\rm F}} + R_{\mathrm{pp}}(Q) \bigg]^{-1} 
\label{eq:Gamma_Q}  
\end{equation}
expressed in terms of the renormalized particle-particle bubble $R_{\mathrm{pp}}(Q)$ 
\begin{equation}
R_{\mathrm{pp}}(Q) = \!\! \int \!\!\!\frac{d\mathbf{k}}{(2\pi)^3} \bigg[ \int \frac{d \omega}{2 \pi} G_\sigma(k) G_{\bar{\sigma}} (Q-k) - \frac{m}{\mathbf{k}^2} \! \bigg] .
\label{eq:Rpp_Q}
\end{equation}
Equations (\ref{eq:G_k})-(\ref{eq:Rpp_Q}) have to be solved self-consistently by a numerical procedure that relies on Fourier transforming back and forth between the $(\mathbf{k},i\omega)$ or $(\mathbf{Q},i\Omega)$ spaces and the $(\mathbf{r},\tau)$ space (see Refs.~\cite{Haussmann-1994,Haussmann-2007,FrankThesis-2018,PPS-2019} for a description of the implementation of this numerical procedure at finite temperature, 
and Appendix~\ref{app:details_numerical_T=0} for additional details on the implementation at zero temperature).  
Note that, owing to rotational invariance, $G_{\sigma}(\mathbf{k},i \omega) = G_{\sigma}(|\mathbf{k}|,i \omega)$ and $\Gamma({\mathbf{Q},i \Omega})=\Gamma(|{\mathbf{Q}|,i \Omega})$.
In addition, for given densities $n_\sigma$, the chemical potentials $\mu_\sigma$ in Eqs.~\mbox{(\ref{eq:G_k})-(\ref{eq:Rpp_Q})} are determined by inverting the density equations
\begin{equation}
n_\sigma=\int\!\! \frac{d {\bf k}}{(2\pi)^3}  \frac{d \omega}{2\pi}\, e^{i  \omega 0^+}  \, G_\sigma(k) \, .
\label{eq:density}
\end{equation}

Throughout this article, the spin polarization $p=(n_\uparrow-n_\downarrow)/(n_\uparrow+n_\downarrow)$ (such that $p>0$ for a majority of spin-up fermions considered here without loss of generality) will act as the 
``tuning parameter'' that drives the evolution of the Fermi gas for given inter-particle coupling. 
Accordingly, the dimensionless coupling $(k_{\rm F} a_{\rm F})^{-1}$ will be an additional tuning parameter, which is here expressed in terms of the (effective) Fermi wave vector $k_{\rm F}=(3 \pi^2 n)^{1/3}$ where $n=n_\uparrow+n_\downarrow$ is the total density. With this definition, $k_{\rm F}$ corresponds to the non-interacting Fermi wave vector of an unpolarized system with the same density $n$. 
An alternative choice, often utilized in the context of ultra-cold gases (although not employed here), prefers instead the dimensionless coupling  $(k_{{\rm F}\uparrow} a_{\rm F})^{-1}$ where  $k_{{\rm F} \uparrow}=(6 \pi^2 n_\uparrow)^{1/3}$ is the non-interacting Fermi wave vector of the majority species. 


\section{Luttinger theorem}
\label{sec:Luttinger}
A key feature of the normal phase of a polarized Fermi gas at zero temperature is the validity of the Luttinger theorem \cite{Luttinger-1960}. 
For a partially polarized Fermi liquid here considered, the theorem states that the volume enclosed by the Fermi surfaces of each of the two spin components remains the same as for a non-interacting gas \cite{Pieri-2017}. 
In the isotropic case, this implies that the radii of the two Fermi spheres are not affected by the inter-particle interaction, thereby remaining equal to the non-interacting Fermi wave vectors 
$k_{{\rm F} \sigma}=(6 \pi^2 n_\sigma)^{1/3}$. 

This theorem was originally proved for the exact theory, and it is not \emph{a priori\/} guaranteed that it holds also for an approximate theory. 
In Ref.~\cite{Pieri-2017} an analytic proof was provided that, for a polarized Fermi gas, a conserving self-consistent approximation (like the $t$-matrix approach considered in Sec.~\ref{sec:theo} above) 
does satisfy the Luttinger theorem. 
This makes the $t$-matrix approach mostly suited for a description of an imbalanced Fermi liquid phase, as it gives full control on the location of the Fermi surfaces for the two spin components. 
This important property, however, is not shared by the non-self-consistent $t$-matrix approximation. 

\begin{figure}[t]    
\includegraphics[angle=0,width=1.01\columnwidth]{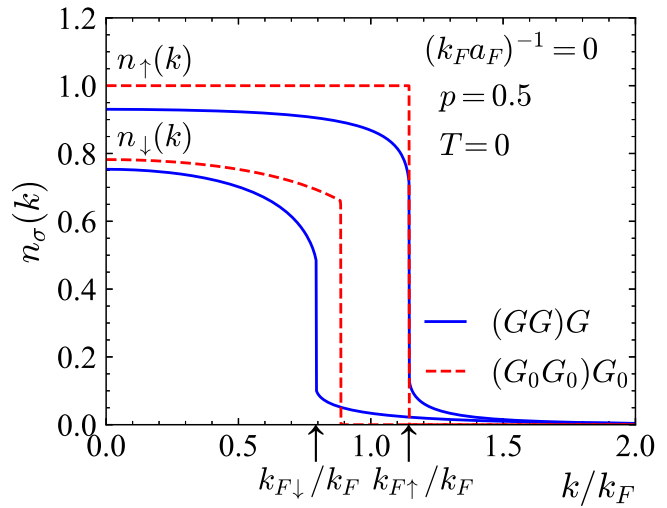}
\caption{Momentum distributions $n_\sigma(\mathbf{k})$ of the two spin components vs $k=|\mathbf{k}|$ [in units of the effective Fermi wave vector $k_{\rm F}=(3 \pi^2 n)^{1/3}$] 
             at unitarity [$(k_{\rm F} a_{\rm F})^{-1}=0$], polarization $p=0.5$, and zero temperature, calculated within both the self-consistent $t$-matrix approach (blue solid lines) 
             and the non-self-consistent $t$-matrix approach (red dashed lines). The arrows on the horizontal axis indicate the Fermi wave vectors $k_{{\rm F} \sigma}=(6 \pi^2 n_\sigma)^{1/3}$ 
             corresponding to the positions of the Fermi surfaces for a non-interacting Fermi gas with spin $\sigma$.}
\label{Figure-1}
\end{figure}

As an example, Fig.~\ref{Figure-1} shows the momentum distributions $n_\sigma(\mathbf{k})=\int_{-\infty}^{+\infty} d\omega \,  e^{i \omega 0^+} G_\sigma (\mathbf{k}, i\omega)$ at unitarity 
[$(k_{\rm F} a_{\rm F})^{-1}=0$] for polarization $p=0.5$, calculated within the fully self-consistent $t$-matrix approach and the non-self-consistent $t$-matrix approach (which corresponds to modifying Eqs.~(\ref{eq:Sigma_k}) and (\ref{eq:Rpp_Q}) by replacing all $G_{\sigma}$ with non-interacting Green's functions $G_{0\sigma}$).
It is evident from this figure that the self-consistent $t$-matrix approach satisfies the Luttinger theorem as it places the Fermi steps exactly at $|\mathbf{k}|=k_{{\rm F} \sigma}=(6 \pi^2 n_\sigma)^{1/3}$ 
(indicated by the arrows on the horizontal axis), while the non-self-consistent $t$-matrix approach violates this condition for the minority component. 
For the majority component, on the other hand, the non-self-consistent $t$-matrix does not violate the theorem to the extent that it incorrectly treats the majority atoms as completely non-interacting. 
This pathological behavior, which was first pointed out in Ref.~\cite{Schneider-2009}, occurs whenever the chemical potential of the minority component $\mu_\downarrow$ is negative 
(an analytical proof of this statement is given in Appendix~\ref{app:failure_NSC}).
This represents a further shortcoming of the non-self-consistent approach for the polarized Fermi gas which is overcome by the fully self-consistent one.

Alternative $t$-matrix approaches based on truncating the Dyson equation (\ref{eq:G_k}) at first order in $\Sigma_\sigma(k)$ (like in the original approach by Nozi{\`e}res and Schmitt-Rink \cite{Nozieres-1985}) 
are also consistent with the Luttinger theorem, but they incur in other unphysical results like negative values of the quasi-particle residues at intermediate polarizations \cite{Urban-2014,Durel-2020}. 
By retaining the full Dyson equation (\ref{eq:G_k}) (like in our approach), one can overcome the problem of the negative quasi-particle residues; yet, the Luttinger theorem is usually violated if full self-consistency is not 
duly implemented \cite{Durel-2020}.

Finally, we mention that the validity of the Luttinger theorem is equivalent to the following condition for the chemical potentials:
\begin{equation}
\mu_\sigma=\frac{k_{F\sigma}^{2}}{2 m} + \Sigma_\sigma(|\mathbf{k}|=k_{{\rm F} \sigma}, \, i\omega=0) \, .
\label{eq:mu_sigma}
\end{equation}
In practice, in our numerical calculations we have enforced Eq.~(\ref{eq:mu_sigma}) and explicitly verified that Eq.~(\ref{eq:density}) for the densities is always satisfied (within a relative error at most of $0.05\%$).

\begin{figure}[h]  
\includegraphics[angle=0,width=0.93\columnwidth]{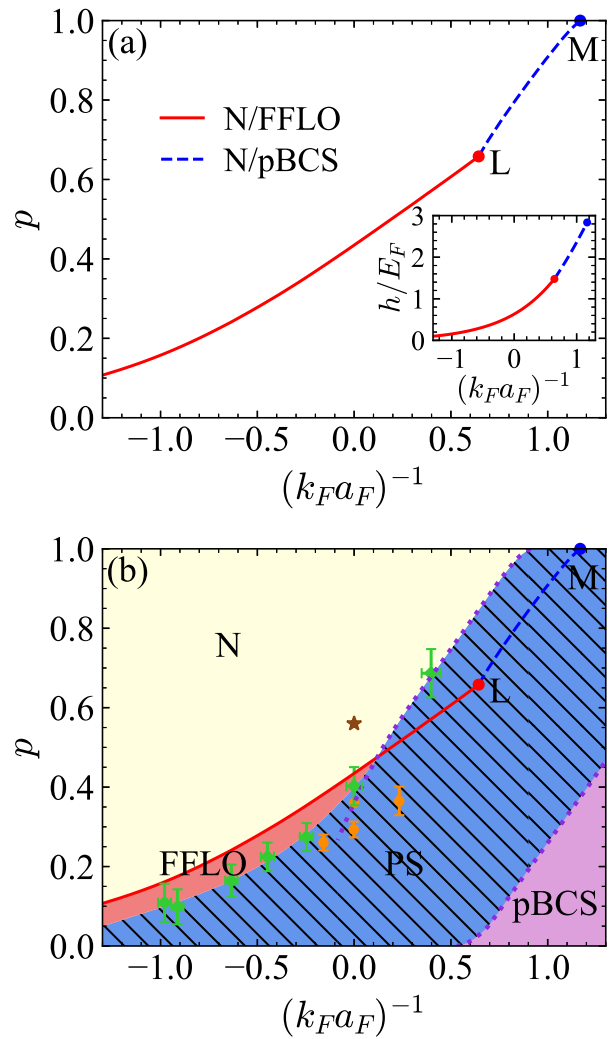}
\caption{Polarization-vs-coupling phase diagram at zero temperature. 
             (a) Phase diagram obtained by the self-consistent $t$-matrix approach for the second-order phase transition between the normal phase (N) and either an FFLO superfluid (full line) 
              or a polarized BCS (pBCS) superfluid (dashed line). At the Lifshitz point (L) the instability changes from FFLO to polarized BCS superfluid (pBCS). 
             The end point of the transition line at $p=1$ corresponds to the polaron-to-molecule transition point M.
             Inset: Corresponding critical field $h = (\mu_{\uparrow}-\mu_{\downarrow})/2$ vs coupling. 
             (b) Phase diagram taking into account the phase separation (hatched PS region) between the normal (N) and the pBCS superfluid phases, as obtained from experimental 
             works [Refs.~\cite{Shin-2008-PRL,Shin-2008-Nature} (circles) and Ref.~\cite{Olsen-2015} (diamonds)] and from QMC calculations \cite{Pilati-2008} (dotted lines). 
             The star represents the result from Ref.~\cite{Bulgac-2008} for the N/FFLO phase transition at unitarity. 
             \label{Figure-2}}
\end{figure}


\section{Zero-temperature phase diagram} 
\label{sec:phase_diagram}
In this Section we consider the description of the zero-temperature phase diagram for the polarized Fermi gas in terms of the self-consistent $t$-matrix approach. 

Accordingly, the critical polarization $p_c$ for the second-order normal-to-superfluid transition is determined by monitoring the momentum dependence of the pair propagator (\ref{eq:Gamma_Q}) at zero frequency. 
The transition point corresponds to the divergence of $\Gamma(|\mathbf{Q}|, 0)$ at some value $Q_0$ of $|\mathbf{Q}|$, signalling the divergence of the pairing susceptibility $\chi_{\rm pair}(|{\bf Q}|)$ at the pair wave vector $Q_0$ \cite{Pini-PRR-2021}. 
The corresponding condition
\begin{equation}
\big[\Gamma(|\mathbf{Q}|=Q_0,i\Omega=0)|_{p=p_c}\big]^{-1}=0 
\label{eq:Thouless}
\end{equation}
generalizes the Thouless criterion \cite{Thouless-1960} to situations when the superfluid phase is of the FFLO type. 
Specifically, when $Q_0=0$ the transition is toward a polarized BCS superfluid (pBCS), while when $Q_0 \neq 0$ the transition is toward an FFLO superfluid.
Note that, at finite temperature, the condition (\ref{eq:Thouless}) for $Q_0 \neq 0$ would lead to a diverging self-energy (\ref{eq:Sigma_k}) for all frequencies and momenta (see Refs.~\cite{Shimahara-1998,Shimahara-1999,Ohashi-2002,Jakubczyk-2017,Wang-2018,Zdybel-2021} for a discussion of this problem in related approaches). 
In practice, what occurs is that the condition for the FFLO transition is never \emph{exactly\/} satisfied at finite temperature, thereby confining a truly FFLO phase to $T=0$ while leaving the system in the normal phase for any $T\neq 0$ 
(albeit with the presence of strong FFLO fluctuations) \cite{Pini-PRR-2021}. 
It is for this reason that, in order to precisely identify the occurrence of the quantum phase transition, it is important to implement the numerical calculations \emph{exactly\/} at $T=0$.
Any extrapolation in terms of finite-temperature results would, in fact, lead to large uncertainties in the location of the FFLO critical line. 

\begin{figure}[h]  
\includegraphics[angle=0,width=1.0\columnwidth]{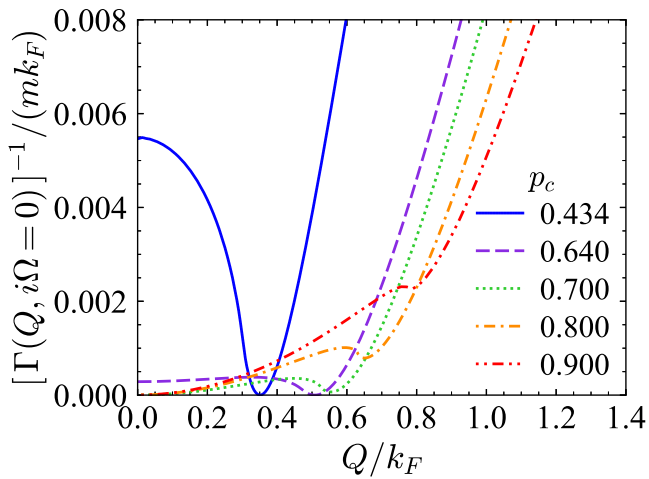}
\caption{Momentum dependence of the inverse of the pair-propagator $\Gamma(|{\bf Q}|,i\Omega=0)^{-1}$ (in units of $m k_{\rm F}$) for zero frequency at the critical polarization $p_c$, 
             for different values of $p_c$ that correspond to $(k_{\rm F} a_{\rm F})^{-1} = (0.00, 0.59, 0.69, 0.83, 0.99)$, from top to bottom.}
\label{Figure-3}
\end{figure}

Figure~\ref{Figure-2}(a) shows the critical polarization vs coupling obtained at zero temperature from the generalized Thouless criterion (\ref{eq:Thouless}) within the present self-consistent $t$-matrix approach. 
A special role in this panel (as well as in panel (b) below) is played by the Lifshitz point L, which is quite generally identified as the point where the disordered phase meets two phases with uniform and modulated order, 
respectively \cite{Chaikin-Lubenski-1995}.
In the present case, this is where the normal phase meets two superfluid phases, one with spatial uniform order and the other one with spatially modulated order \cite{Son-2006}.
Our calculations locate the Lifshitz point at $((k_{\rm F}a_{\rm F})^{-1}_{\rm L},p_{\rm L})=(0.643,0.658)$, which correspond to notably larger coupling and smaller polarization with respect to the mean-field results 
$((k_{\rm F}a_{\rm F})^{-1}_{\rm L},p_{\rm L})=(0.26,0.98)$ \cite{Physics-Reports-2018}. 
On the left side of the Lifshitz point L in Fig.~\ref{Figure-2}(a), the momentum $Q_0$ at the transition is different from zero 
and the phase boundary separates the normal phase (N) from the FFLO superfluid phase. 
On the right side of the Lifshitz point in Fig.~\ref{Figure-2}(a), on the other hand, the momentum $Q_0$ at the transition vanishes and the phase boundary separates the normal phase from a polarized BCS superfluid phase (pBCS). 
The phase boundary terminates at the polaron-to-molecule transition point M with $(k_{\rm F}a_{\rm F})^{-1}_\text{M}=1.17$ and $p=1$. 
This position for M is consistent (within extrapolation errors) with previous calculations using similar self-consistent $t$-matrix approaches in the polaronic limit~\cite{Hu-2018,Frank-2018}, 
and is rather close to the diagrammatic Monte Carlo value $(k_{\rm F}a_{\rm F})^{-1}_\text{M}=1.14(2)$ \cite{Prokofev-2008}. 
 
We have further found that the vanishing of the pair momentum $Q_0$ at the Lifshitz point occurs with a sudden jump from a value $Q_0 \neq 0$, as it can be seen from the behavior of 
$\Gamma(|{\bf Q}|, i\omega=0)^{-1}$ shown in Fig.~\ref{Figure-3}. 
One sees from this figure that, upon increasing the coupling strength (and thus $p_c$), the local maximum of $\Gamma(|{\bf Q}|, i\omega=0)^{-1}$ at $Q_0=0$ turns into a local minimum close to the L point, 
which eventually becomes the absolute minimum as the L point is crossed.  

Our calculations did not consider the occurrence of phase separation, since accounting for this possibility would require us to extend our $t$-matrix approach to the superfluid phase. 
Previous experimental \cite{Shin-2008-PRL,Shin-2008-Nature,Olsen-2015} and quantum Monte Carlo studies \cite{Pilati-2008} have pointed to the existence of a rather broad phase separation region in the phase diagram. Accordingly, Fig.~\ref{Figure-2}(b) shows the expected region of phase separation (PS), obtained by interpolating experimental \cite{Shin-2008-PRL,Shin-2008-Nature,Olsen-2015} and quantum Monte Carlo data \cite{Pilati-2008}. 
By comparing this region with our second-order phase transition lines, and assuming that (as it occurs in the weak-coupling limit \cite{Takada-1969}) the boundary of the phase separation region with the normal phase essentially coincides with the boundary of the phase separation region with the FFLO phase, we conclude that the FFLO phase should still be present in the region of Fig.~\ref{Figure-2}(b) evidenced in red, which extends from weak coupling up to $(k_{\rm F} a_{\rm F})^{-1}\simeq 0.1$ (albeit in a rather narrow range of polarization). 
Note that, for polarization below this region, phase separation should occur between a standard spin-balanced BCS superfluid and an FFLO polarized superfluid. 

The presence of such a narrow FFLO region as shown in Fig.~\ref{Figure-2}(b) has probably escaped experimental detection \cite{Shin-2008-PRL,Shin-2008-Nature,Olsen-2015}, because in a harmonic trap the FFLO superfluid is confined in a narrow shell surrounding the unpolarized BCS core. 
It is then hard to distinguish this narrow FFLO shell from the (significantly larger) surrounding shell made of a normal polarized Fermi gas. 
In this context, experiments using a box-like trap \cite{Mukherjee-2017,Hueck-2018,Shkredov-2021,Navon-2021,Navon-2022} should be able to avoid this problem.  
The calculation of Ref.~\cite{Pilati-2008}, on the other hand, excluded from the outset an FFLO solution, due to to the choice of the trial wave-function used in the fixed-node diffusion QMC simulations.

We note further from Fig.~\ref{Figure-2}(b) that, at unitarity, the second-order FFLO phase transition is not covered by phase separation.
This is in contrast to what is found within mean field \cite{Sheehy-2007}, but is in line with the prediction of a density functional theory with input from quantum Monte Carlo data \cite{Bulgac-2008} 
[star in Fig~\ref{Figure-2}(b)] although with the different value $p_c=0.56$ for the critical polarization (recall that $p_c=0.434$ in our calculation). 
In terms of the average chemical potential $\mu= (\mu_\uparrow + \mu_\downarrow)/2$ and of the Zeeman splitting field $h= (\mu_\uparrow - \mu_\downarrow)/2$, at unitarity we find that $(\mu/h)_c=1.53$ at the FFLO phase transition. 
This value can be compared with the (less precise) value $(\mu/h)_c=1.28 \pm 0.15$, obtained in Ref.~\cite{Frank-2018} by extrapolating to $T\to 0$ the results of the self-consistent $t$-matrix approach at finite-temperature.

For couplings $(k_{\rm F} a_{\rm F})^{-1} \gtrsim 0.1$, on the other hand, the FFLO phase, the Lifshitz point, as well as the transition line connecting the L and M points, are hidden under the region of 
phase separation (PS).
Nevertheless, we have verified that the compressibility matrix remains positive-definite for all polarizations $p \ge p_c $ to which we have access, indicating that the normal phase is mechanically stable even in the region of phase separation above $p_c$, where it could still be present as a metastable phase.
As a consequence, the second-order transition line and the L point in Fig.~\ref{Figure-2}(b) could, in principle, be reached even in this region along this metastable phase. 
On the strong-coupling side of the phase diagram, phase separation eventually gives way to a polarized BCS superfluid (the pBCS region in the right-bottom corner of Fig.~\ref{Figure-2}(b)). 

\begin{figure}[t]  
\includegraphics[angle=0,width=1.0\columnwidth]{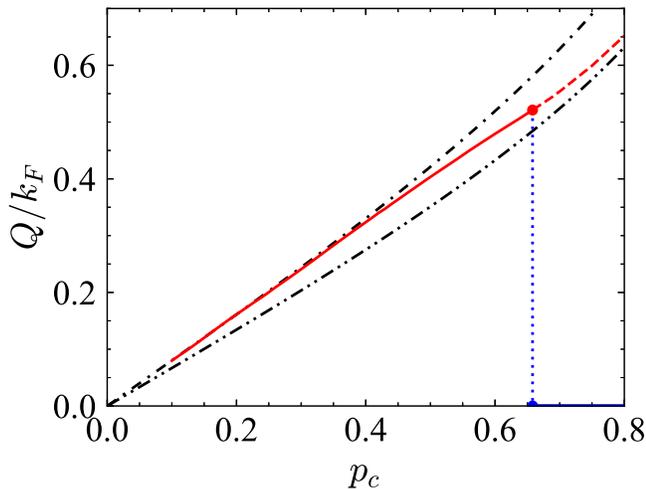}
\caption{Pair wave vector $Q_0$ (in units of $k_{\rm F}$) vs critical polarization $p_c$ at the superfluid phase transition. 
             At the Lifshitz point where $p_c=p_{\rm L}=0.658$, $Q_0$ jumps abruptly from $Q_0 =0.521 k_{\rm F}=1.076 |k_{{\rm F} \uparrow}-k_{{\rm F} \downarrow}|$ to $Q_0=0$ (vertical dotted line). 
             The dashed line for $p_c>p_{\rm L}$ corresponds to the secondary minimum of $\Gamma(\mathbf{Q},i\Omega=0)^{-1}$ (the absolute minimum is at $Q_0=0$). 
             The dash-dotted line represents the Fulde-Ferrell mean-field result $Q_0=1.2 |k_{{\rm F} \uparrow}-k_{{\rm F} \downarrow}|$ \cite{Takada-1969}, while the dashed-double-dotted line corresponds to  
             $Q = |k_{{\rm F} \uparrow}-k_{{\rm F} \downarrow}|$.}
\label{Figure-4}
\end{figure}

Finally, Fig.~\ref{Figure-4} shows the pair wave vector $Q_0$ at the transition vs the critical polarization $p_c$ along the N/FFLO (red full line) and the N/pBCS (blue dashed line) phase boundaries 
of Fig.~\ref{Figure-2}(a). 
The red dashed line of Fig.~\ref{Figure-4} corresponds instead to a secondary local minimum of $\Gamma(\mathbf{Q},i\Omega=0)^{-1}$, as its absolute minimum jumps abruptly from $Q_0 =0.521 k_{\rm F}=1.076 |k_{{\rm F} \uparrow}-k_{{\rm F} \downarrow}|$ to $Q_0=0$ at the Lifshitz point $p_c=p_{\rm L}$  (vertical dotted line), as mentioned above. 
The dash-dotted line in Fig.~\ref{Figure-4} corresponds to the Fulde-Ferrell mean-field result $Q_0=1.2 |k_{{\rm F} \uparrow}-k_{{\rm F} \downarrow}|$ \cite{Takada-1969}, which is correctly recovered for weak coupling (corresponding to $p_c \to 0$).
Note also that, in the 3D system we are considering, the pair wave vector $Q_0$ remains always larger than $|k_{{\rm F} \uparrow}-k_{{\rm F} \downarrow}|$ (dashed-double-dotted line), which is the expected result in 1D. 


\section{Evolution of the Fermi liquid phase with polarization}
\label{sec:Evolution_FL}
In this Section, we characterize the evolution with polarization of the Fermi liquid phase at zero temperature from the polaronic limit ($p=1$) to the superfluid QCP ($p=p_c$) for various couplings $(k_{\rm F} a_{\rm F})^{-1}$. 
In particular, in Sec.~\ref{subsec:residue_effmass} we present this evolution in terms of the quasi-particle residues and effective masses, while in Sec.~\ref{subsec:dynamical_scaling} we analyze the evolution 
of the self-energies with polarization and consider the scaling behavior of dynamical quantities close to the FFLO QCP.

\subsection{Quasi-particle residues and effective masses}
\label{subsec:residue_effmass}
By assuming a Fermi liquid behavior  \cite{Nozieres-1964}, the quasi-particle residue $Z_\sigma$ and the effective mass $m^*_\sigma$ of quasi particles with spin component $\sigma$ are obtained by \cite{footnote-Z-from-imaginary-part}: 
\begin{align}
\label{eq:residue}
Z_\sigma &= \bigg[ 1- \frac{\partial \, \text{Im}  \Sigma_\sigma (k_{{\rm F} \sigma},i\omega)}{ \partial \omega} \Big|_{\omega=0^+} \bigg]^{-1} \\
\label{eq:effective_mass}
\frac{m}{m^*_\sigma} &= Z_\sigma \bigg[1 +\frac{m}{k_{{\rm F} \sigma}} \frac{\partial  \text{Re} \Sigma_\sigma (\mathbf{k},i 0^+)}{\partial |\mathbf{k}|} \Big|_{|\mathbf{k}|=k_{{\rm F} \! \sigma}} \bigg] \, .
\end{align}
Note that in Eq.~(\ref{eq:residue}) we have used the analytic properties of $\Sigma_\sigma (k_{{\rm F} \sigma},\zeta)$ in the upper-half complex plane of $\zeta$ to calculate the derivative at $\zeta=0+i0^+$  along the imaginary frequency axis rather than on the real frequency axis, avoiding in this way the need of analytic continuation to the real frequency axis (see also Refs.~\cite{Prokofev-2008b,Hu-2018,Arsenault-2012,Schafer-2021}). 

\begin{figure}[tb]   
\includegraphics[angle=0,width=1.0\columnwidth]{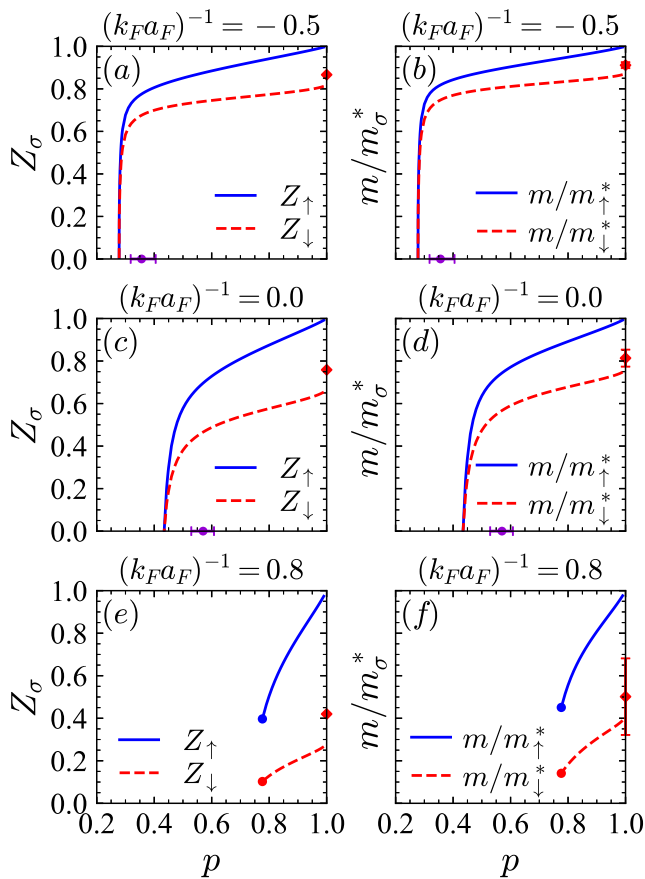}
\caption{Quasi-particle residues $Z_\sigma$ and inverse effective mass ratios $m/m^*_\sigma$ for the two spin components $\sigma=(\uparrow,\downarrow)$ as functions of polarization $p$, at zero temperature 
             and for the coupling values $(k_{\rm F} a_{\rm F})^{-1}=(-0.5,0.0,0.8)$. 
              The violet circles on the horizontal axis in panels (a) to (d) indicate the value  of the crossover polarization  $p^*$ that sets the boundary of the strong FFLO fluctuation region near the FFLO QCP. Red diamonds: diagrammatic Monte Carlo results \cite{Vlietinck-2013} (data outside unitarity have been interpolated).}
\label{Figure-5}
\end{figure}

Figure~\ref{Figure-5} shows the evolution of the two quantities (\ref{eq:residue}) and (\ref{eq:effective_mass}) in the Fermi liquid normal phase as functions of polarization $p$, for three characteristic couplings 
$(k_{\rm F} a_{\rm F})^{-1} =(-0.5,0.0,0.8)$. 

Even in the polaronic limit ($p=1^{-}$), the spin-dependence of the results stems from the spin-dependence of the self-energy for the spin-imbalanced system we are considering. In fact, in this limit the majority component is completely non-interacting, so that $Z_\uparrow=1$ and $m/m^*_\uparrow=1$. 
The minority component is instead dressed by the majority atoms through the attractive inter-particle interaction, and forms a quasi-particle (an attractive polaron) with $Z_\downarrow<1$ and $m/m^*_\downarrow<1$ 
even when $p=1$.  For comparison, the values of the residues and effective masses for the minority component obtained in the polaronic limit by the diagrammatic Monte Carlo method \cite{Vlietinck-2013} are also reported in Fig.~\ref{Figure-5} (red diamonds).

 Upon decreasing $p$, the effects of the attractive interaction become more evident for both spin components, and the values of $Z_\sigma$ and $m/m^*_\sigma$ decrease monotonically. 
At the QCP, however, the nature of the incipient superfluid phase strongly influences the behavior of the quasi-particles.  
When the transition is to an FFLO superfluid [Figs.~\ref{Figure-5}(a)$-$(d)], the quasi-particle residues $Z_\sigma$ vanish and the effective masses $m^*_\sigma$ diverge at the QCP. 
When the transition is instead to the more standard BCS polarized superfluid [Figs.~\ref{Figure-5}(e)$-$(f)], $Z_\sigma$ and $m/m^*_\sigma$ remain finite although their values get strongly reduced with respect to the polaronic limit. 

In the case of an FFLO QCP, one can readily identify two different regimes of polarization, namely,
a regime at large polarization where the residues decrease almost linearly by decreasing $p$, and a regime close to the FFLO QCP where the decrease with $p$ becomes strongly non-linear. 
It is evident from Fig.~\ref{Figure-5} that the non-linear region increases its extension as the coupling gets stronger (as long as an FFLO QCP is present). 
The boundary of this region can be determined by fitting the linear behavior occurring in the large-$p$ regime, and by finding the value $p^*$ of the polarization at which the residues $Z_\sigma$ deviate 
more than a certain percentage (which we have conventionally fixed at $5\%$) from the fitted line.  An error bar for $p^*$ is also included, corresponding to polarizations at which the deviation from the fitted line spans from $2.5\%$ to $10\%$ (for more details about the fitting procedure see Appendix~\ref{app:details_numerical_T=0}). In Figs.~\ref{Figure-5}(a)$-$(d) the values of $p^*$ are identified by the circles with error bars on the horizontal axis. One can identify $p^*$ as a crossover polarization below which strong FFLO fluctuations set in as a precursor of the FFLO QCP.

Our further finding of strongly renormalized (yet ultimately well-defined) quasi-particles when the QCP is toward a BCS polarized superfluid [Figs.~\ref{Figure-5}(e)$-$(f)] is in line with what was predicted in Ref.~\cite{Strack-2014} for a corresponding QCP in 2D. 
Nevertheless, the vanishing of the quasi-particle residues and the divergence of the effective masses at an FFLO QCP [Figs.~\ref{Figure-5}(a)$-$(d)] signal a breakdown of the quasi-particle description 
of the Fermi liquid phase at the QCP. 
This finding is in line with what was predicted for the FFLO QCP in 2D within an $\epsilon$-expansion approach \cite{James-2010}. 
More generally, analogous breakdowns of the Fermi liquid theory are known to occur at QCPs toward phases with periodic modulations, like for antiferromagnetic QCPs in heavy-fermion materials \cite{Varma-2002,Lohneysen-2007,Senthil-2008}.

The divergence of the effective masses $m^*_\sigma$ reported in Figs.~\ref{Figure-5}(a)$-$(d) is a consequence of the vanishing of the quasi-particle residues $Z_\sigma$ through the relationship (\ref{eq:effective_mass}). 
We have indeed verified that the only singular contribution to $m^*_\sigma$ originates from $Z_\sigma$, while the term related to $\partial \text{Re} \Sigma_\sigma (\mathbf{k},i\omega=0^+)/\partial |\mathbf{k}|$
is regular and thus contributes only a non-singular multiplicative term. 

\subsection{Self-energies close to the FFLO QCP and dynamical scaling}
\label{subsec:dynamical_scaling}

\begin{figure}[tb]   
\includegraphics[angle=0,width=1.0\columnwidth]{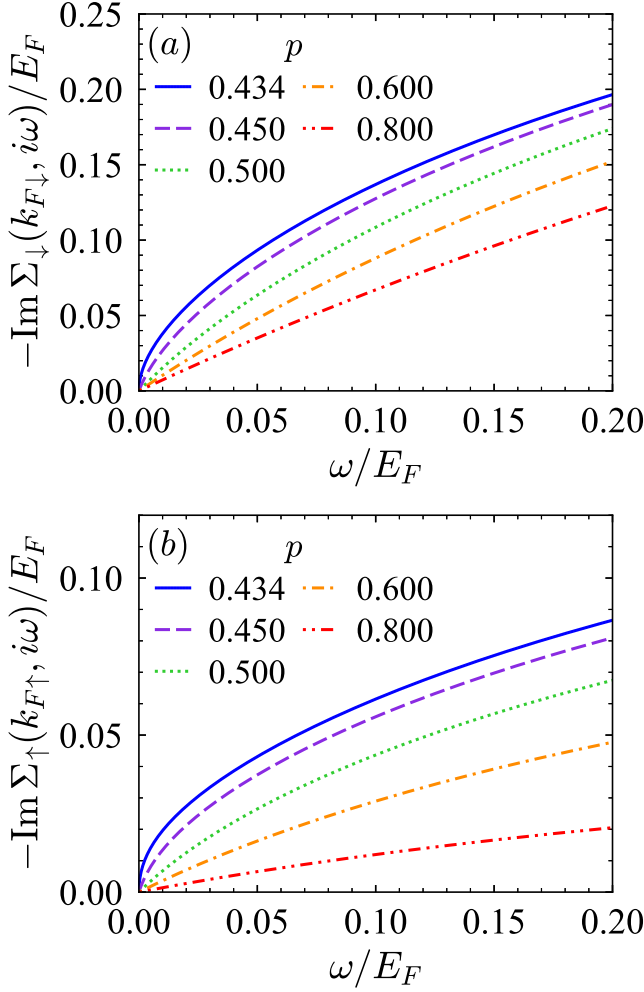}
\caption{(Minus the) imaginary part $-\text{Im} \Sigma_\sigma(k_{{\rm F} \sigma},i\omega)$ of self-energies [in units of $E_{\rm F}=k_{\rm F}^2/(2m)$] at the Fermi surface for both spin components  
              vs the frequency $\omega$ along the imaginary axis, for various polarizations at zero temperature and unitarity $[(k_{\rm F} a_{\rm F})^{-1}=0]$. 
              The value $p=0.434$ corresponds to the critical polarization $p_c$ at unitarity.}
\label{Figure-6}
\end{figure}

The vanishing of the quasi-particle residues $Z_\sigma$ at the  FFLO QCP requires the derivative  of $ \text{Im}  \Sigma_\sigma (k_{{\rm F} \sigma},i\omega)$ with respect to $\omega$ to diverge at $\omega=0$ 
[cf.~Eq. (\ref{eq:residue})]. 
It is interesting to verify numerically how this non-Fermi liquid behavior is attained in practice upon approaching the QCP by progressively decreasing the polarization.  

\begin{figure}[tb]  
\includegraphics[angle=0,width=1.0\columnwidth]{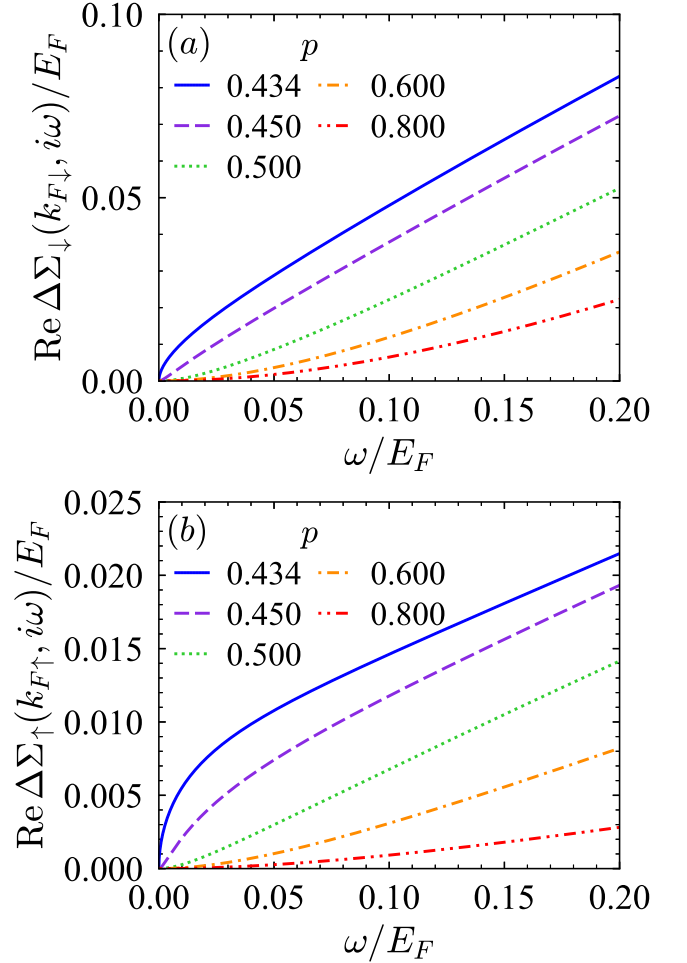}
\caption{Real part of $\Delta \Sigma_\sigma(k_{{\rm F} \sigma}, i\omega)\equiv \Sigma_\sigma(k_{{\rm F} \sigma}, i\omega) -\Sigma_\sigma(k_{{\rm F} \sigma}, 0)$  [in units of $E_{\rm F}=k_{\rm F}^2/(2m)$] 
              at the Fermi surface  for both spin components vs the frequency $\omega$ along the imaginary axis, for various polarizations at zero temperature and unitarity $[(k_{\rm F} a_{\rm F})^{-1}=0]$.
              The value $p=0.434$ corresponds to the critical polarization $p_c$ at unitarity.}
\label{Figure-7}
\end{figure}

Figure~\ref{Figure-6} shows the evolution of $-\text{Im} \Sigma_\sigma (k_{{\rm F} \sigma},i\omega)$ with polarization at unitarity for both spin components.
Note how the linear behavior of $- \text{Im} \Sigma_\sigma(k_{{\rm F} \sigma}, i\omega)=(1-1/Z_\sigma) \, \omega$ for $\omega \simeq 0$ characteristic of a Fermi liquid is gradually replaced by a square-root behavior
for decreasing polarization.
Specifically, the linear region progressively reduces its extension to smaller $\omega$ until it shrinks to zero at the QCP, where $- \text{Im} \Sigma_\sigma(k_{{\rm F} \sigma}, i\omega) \sim \omega^{1/2}$.
In particular, at low $\omega$ we obtain a robust fit to the numerical data for $-\text{Im} \Sigma_\sigma(k_{{\rm F} \sigma}, i\omega)$ with the function $C_\sigma \omega^{1/2}$ 
(for instance, for the data of Fig.~\ref{Figure-6} at the QCP and $\omega < 0.02 E_{\rm F}$ we obtain a value $R^2>0.9995$ for the coefficient of determination associated with the fit to both components
\cite{reference-R^2}).
By repeating the same kind of fit for different couplings (but still remaining at the FFLO QCP), we always find a similar square-root behavior for $ -\text{Im}  \Sigma_\sigma (k_{{\rm F} \sigma},i\omega)$.
This behavior is also shared by $\text{Re} \Delta \Sigma_\sigma(k_{{\rm F} \sigma}, i\omega)\equiv \text{Re} [\Sigma_\sigma(k_{{\rm F} \sigma}, i\omega) -\Sigma_\sigma(k_{{\rm F} \sigma}, 0)]$,
as shown in Fig.~\ref{Figure-7}.
On the other hand, along the N/pBCS critical line of Fig.~\ref{Figure-2}(a), $- {\rm Im} \Sigma_\sigma(k_{{\rm F} \sigma}, i\omega)$ remains linear in $\omega$, 
consistently with the finite value that we have found for the residue in this case.

It is further interesting to note that the square-root behavior, that we have found for both $-\text{Im} \Sigma_\sigma(k_{{\rm F} \sigma}, i\omega)$ and $\Delta\text{Re} \Sigma_\sigma(k_{{\rm F} \sigma}, i\omega)$,
can be related to the dynamical critical exponent $z$ of the dynamic pairing susceptibility. 
To make this connection explicit, it is sufficient to analyze the small-frequency behavior of the retarded self-energies $\Sigma^{\text{R}}_\sigma(k_{{\rm F} \sigma}, \tilde{\omega})$, which is obtained from 
$\Sigma_\sigma(k_{{\rm F} \sigma}, i\omega)$ after analytic continuation $i\omega \to \tilde{\omega}+i 0^+$ in the complex upper-half plane (where the tilde signifies that the frequency is taken along the real axis).
To this end, we begin by considering the small-frequency behavior as obtained from the fitting procedure discussed above on the positive imaginary frequency axis:
\begin{eqnarray}
\text{Re} \Sigma_\sigma(k_{{\rm F} \sigma}, i\omega) &=& \Sigma_\sigma(k_{{\rm F} \sigma}, 0) + B_{\sigma}\omega^{1/2}\\
\text{Im} \Sigma_\sigma(k_{{\rm F} \sigma}, i\omega) &=& - C_{\sigma} \omega^{1/2}
\end{eqnarray}
where $C_{\sigma} > 0$ and $B_{\sigma}\ >0$ are real constants.  
Denoting by $\zeta$ the complex frequency in the upper-half plane, for $\zeta \to 0$ these equations imply that
\begin{equation}
\Sigma_\sigma(k_{{\rm F} \sigma}, \zeta) =  \Sigma_\sigma(k_{{\rm F} \sigma}, 0)  + (B_{\sigma} - i C_{\sigma}) \sqrt{- i \zeta} \, ,
\end{equation}
which yields after analytic continuation $\zeta=\tilde{\omega} + i 0^+$ to the real frequency axis:
\begin{eqnarray}
\label{eq-sigmaR}
\Sigma^{\text{R}}_\sigma(k_{{\rm F} \sigma}, \tilde{\omega}) &=& \Sigma_\sigma(k_{{\rm F} \sigma}, 0) + [B_{\sigma} - C_{\sigma}\text{sgn}(\tilde{\omega})] \sqrt{|\tilde{\omega}|/2}
\nonumber\\
&-& i [C_{\sigma}+B_{\sigma}\text{sgn}(\tilde{\omega})] \sqrt{|\tilde{\omega}|/2}
\end{eqnarray}
for $\tilde{\omega} \to 0$. 
Note that the condition $\text{Im} \Sigma^{\rm R} \le 0$ requires $B_{\sigma} \le C_{\sigma}$, which is always satisfied by our fittings.

Equation (\ref{eq-sigmaR}) together with the condition (\ref{eq:mu_sigma}), in turn, yield for the single-particle spectral functions $A_\sigma({\bf k},\tilde{\omega})\equiv -(1/\pi) {\rm Im} G^{\rm R}({\bf k},\tilde{\omega})$  
at $|{\bf k}| = k_{{\rm F} \sigma}$
\begin{equation}
A_\sigma(k_{{\rm F} \sigma},\tilde{\omega}) =\frac{D_{\sigma\pm}}{|\tilde{\omega}|^{1/2}} \, ,
\label{eq:A_omega}
\end{equation}
where the coefficient $D_{\sigma\pm}$ (that depends on the sign of $\tilde{\omega}$) is given by
$$
D_{\sigma\pm} = \frac{1}{\sqrt{2}\pi}\frac{C_{\sigma}+B_{\sigma}\text{sgn}(\tilde{\omega})}{B_\sigma^2+C_\sigma^2} \, .
$$ 
This results should be contrasted with the standard Fermi liquid behavior $A_\sigma(k_{{\rm F} \sigma},\tilde{\omega})=Z_{\sigma}\delta(\tilde{\omega})$. 

It was quite generally argued in Ref.~\cite{Senthil-2008} that, whenever the quasi-particle residue vanishes at QCPs, the single-particle spectral function for 
$|{\bf k}|\simeq k_{{\rm F} \sigma}$ and $\tilde{\omega}\simeq 0$ should have the scaling form:
\begin{equation}
A^{\text{(QCP)}}_\sigma(\mathbf{k},\tilde{\omega}) = \frac{c_{0\sigma}}{|\tilde{\omega}|^{d_{\alpha}/z}} F_0 \bigg[ \frac{c_{1\sigma} \tilde{\omega}}{||\mathbf{k}|-k_{{\rm F} \sigma}|^z}\bigg] \, .
\label{eq:A(kF,omega)_scaling}
\end{equation}
Here, $F_0$ is a universal scaling function, $c_{0\sigma}$ and $c_{1\sigma}$ are non-universal coefficients, and $z$ is the dynamical critical exponent (obtained from the dynamical susceptibility of the ordering 
variable associated with the quantum phase transition). 
The exponent $d_{\alpha}$ equals unity whenever (like in the present case) the quasi-particle residue and the inverse effective mass vanish at the QCP with the same behavior 
[i.e., when $\partial \Sigma_\sigma ({\bf k},0)/\partial |{\bf  k}|$ is regular at $k_{{\rm F}\sigma}$, a property which we have verified numerically as already mentioned]. 
By then setting $d_{\alpha} = 1$ and  taking the limit $|\mathbf{k}|\to k_{{\rm F} \sigma}$ in Eq.~(\ref{eq:A(kF,omega)_scaling}), we recover Eq.~(\ref{eq:A_omega}) with 
$D_{\sigma\pm}=c_{0\sigma}F_0(\pm \infty)$. 
We thus conclude that the exponent 1/2 in Eq.~(\ref{eq:A_omega}) is just what is expected when the dynamical critical exponent $z$ equals 2.

To check the validity of the identification $z=2$ directly by our approach, we may consider the dynamical pairing susceptibility $\chi_\text{pair}(\mathbf{Q},i\Omega)$, which in Ref.~\cite{Pini-PRR-2021} 
was shown to coincide with $\Gamma(\mathbf{Q},i\Omega)$ within the self-consistent $t$-matrix approach.
On the positive imaginary frequency axis, for $|\mathbf{Q}|\simeq Q_0$ and $i\Omega \simeq 0$ we then take for this quantity the following form
\begin{equation}
\chi_\text{pair}(\mathbf{Q},i\Omega) \simeq \frac{(m k_{\rm F})^{-1}}{ a + b (|\mathbf{Q}|-Q_0)^2  - (d_1+i d_2)  i \Omega  } \, , 
\label{eq:chi_pair_QCP}
\end{equation}
where the parameters $(a, b, d_1, d_2)$ are all real and positive, with $a$ vanishing at the QCP. 
[In Ref.~\cite{Perali-2002} a similar expansion was utilized for the non-self-consistent pair-propagator $\Gamma_0(\mathbf{Q},i\Omega)$ in the density balanced case with $Q_0=0$]. 
We have numerically verified that, within the present approach, the expansion (\ref{eq:chi_pair_QCP}) remains valid for all couplings we have considered \cite{footnote_anomalous_dim}. 
By performing in Eq.~(\ref{eq:chi_pair_QCP}) the analytic continuation $i \Omega \to \tilde{\Omega} +i0^+$ from the upper-half complex plane to real frequencies $\tilde{\Omega}$, and introducing the coherence length 
$\xi = \sqrt{b/a}$ \cite{Pistolesi-1996}, we obtain the following scaling behavior for the dynamical pairing susceptibility (on the real frequency axis) \cite{Varma-2002,Sondhi-1997}
\begin{equation}
\chi_\text{pair}(\mathbf{Q},\tilde{\Omega}; \xi)=\frac{\xi^2}{m k_{\rm F} b} \Phi_0[(|\mathbf{Q}|-Q_0)\xi,  m_1 \tilde{\Omega} \xi^2] \, ,
\label{eq:chi_pair_QCP_scaling}
\end{equation}
where $m_1=(d_1+i d_2)/b$ is a non-universal complex constant (with dimension of a mass) and $\Phi_0$ is a universal scaling function. 
From this expression, we conclude that the real frequency $\tilde{\Omega}$ scales with the inverse of a coherence time $\xi_\tau \sim \xi^2$, that corresponds to the dynamical critical exponent $z=2$. 


\section{Finite-temperature effects of the FFLO QCP at unitarity}
\label{sec:finite_temperature}
In this Section, we consider how finite temperature affects the system we are dealing with.
Specifically, we shall focus on the unitary regime $(k_{\rm F} a_{\rm F})^{-1}=0$ where, according to our calculations, the QCP is not hidden by a phase-separation region like that seen experimentally 
(cf.~Fig.~\ref{Figure-2}(b)) and, in particular, is of FFLO type. 

Close to the FFLO QCP, we may assume that the temperature enters the scaling behavior (\ref{eq:chi_pair_QCP_scaling}) for the dynamical pairing susceptibility through an additional $\tilde{\Omega}/T$ scaling, such that for $|\mathbf{Q}|\simeq Q_0$ and $\tilde{\Omega} \simeq 0$ we may write
\begin{equation}
\begin{split}
&\chi_\text{pair}(\mathbf{Q},\tilde{\Omega}; \xi, T)=\frac{\xi^2}{m k_{\rm F} b} \\
& \quad \quad \quad \times \Phi\bigg[(|\mathbf{Q}|-Q_0)\xi, \,  m_1  \tilde{\Omega} \xi^2,  \, \frac{\tilde{\Omega}}{2 \pi T} \bigg] 
\end{split}
\label{eq:chi_pair_QCP_scalingT}
\end{equation}
where $\Phi$ is a new universal scaling function. 
Note that the factor $(2 \pi)^{-1}$ in the $\tilde{\Omega}/T$ scaling is due to the spacing $\Delta \Omega_\nu=2 \pi T$ between two successive bosonic Matsubara frequencies, which, in turn, defines the infrared cutoff introduced by the temperature $T$ itself \cite{Sondhi-1997}. 

\begin{figure}[tb] 
\includegraphics[angle=0,width=1.0\columnwidth]{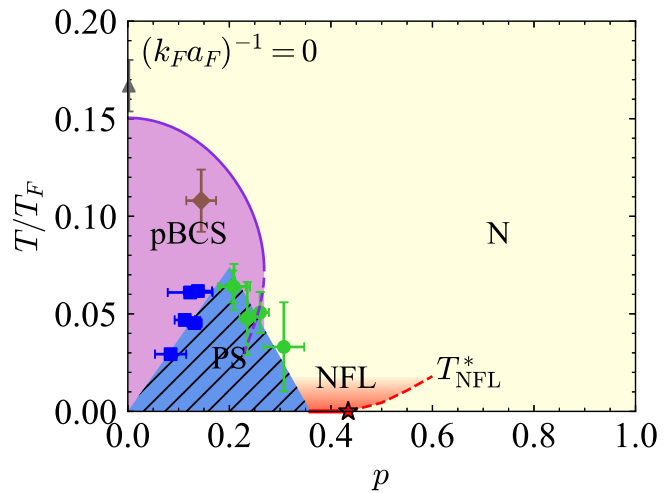}
\caption{Temperature-vs-polarization phase diagram at unitarity.  
              Full line:  Critical temperature $T_c$ [in units of $T_F=E_F=k_{\rm F}^2/(2m)$] for the second-order phase transition between the normal (N) and the polarized BCS superfluid (pBCS) phases obtained by the self-consistent 
              $t$-matrix approach~\cite{Pini-PRA-2021,Pini-PRR-2021}. 
              Triangle and diamond: Experimental data from Refs.~\cite{Sommer-2012} and \cite{Shin-2008-Nature}, respectively, for the second-order superfluid transition. 
              Squares and green circles: Data from Ref.~\cite{Shin-2008-Nature} that delimit the region of phase separation (PS). 
              The red full line at $T=0$ to the right of the PS region corresponds to the FFLO phase, which terminates at the FFLO QCP (red star).   
              Dashed red line: Crossover temperature $T^*_\text{NFL}$ as obtained from Eq.~(\ref{eq:T*_NFL}). 
              Red shaded area: Region of non-Fermi-liquid (NFL) behavior at finite temperature.}
\label{Figure-8}
\end{figure}
 
To identify the temperature region of the non-Fermi liquid behavior above the QCP, one has to compare the two time-scales which appear in Eq.~(\ref{eq:chi_pair_QCP_scalingT}), namely, the thermal time $L_T \sim 1/ T$ that describes the time scale of classical fluctuations, and the coherence time $\xi_\tau \sim \xi^2$ that describes the time scale of quantum fluctuations \cite{Varma-2002,Sondhi-1997}. 
At high temperature, $L_T < \xi_\tau$ and fluctuations are classical.
At low temperature (and for $p> p_c$), on the other hand, $L_T > \xi_\tau$ and fluctuations become quantum such that the system behaves as a Fermi liquid \cite{Varma-2002,Sondhi-1997}. 
The crossover temperature $T^{*}_\text{NFL}$, which separates the Fermi liquid from the non-Fermi-liquid regimes, can thus be determined by setting $L_T \sim \xi_\tau$, that is, $T^{*}_\text{NFL} \sim \xi^{-2}$. 
The constant of proportionality in the definition of $T^*_\text{NFL}$ can further be chosen to comply with the scaling form (\ref{eq:chi_pair_QCP_scalingT}), such that
\begin{equation}
T^*_\text{NFL}=\frac{\xi^{-2}}{2 \pi |m_1 |}  = \frac{a}{2 \pi |d_1+id_2|} \, .
\label{eq:T*_NFL}
\end{equation}
Here, the real parameter $a = a(p,T)$ is obtained from $a = [m k_{\rm F} \, \Gamma(|\mathbf{Q}|=Q_0,i\Omega=0)]^{-1}$ [cf.~Eq.~(\ref{eq:chi_pair_QCP})] for given temperature $T$ and polarization $p$,
while the real parameters $d_1$ and $d_2$ are fixed to their values obtained at the FFLO QCP by fitting $\Gamma(|\mathbf{Q}|=Q_0,i\Omega)$ at small $\Omega$ with the form (\ref{eq:chi_pair_QCP}) (at unitarity, $d_1\simeq 0.040$ and $d_2 \simeq  0.066$). 

The results obtained for $T^*_\text{NFL}$ at unitarity as a function of polarization are shown in Fig.~\ref{Figure-8} (red dashed line). 
To  cast these results in a broader context, Fig.~\ref{Figure-8} reports also additional information on the temperature-vs-polarization phase diagram of the unitary Fermi gas obtained by previous works (as described in detail in the caption of Fig.~\ref{Figure-8}) \cite{footnote-Goulko}.

Note from Fig.~\ref{Figure-8} that, in contrast to the characteristic fan-shaped non-Fermi-liquid region usually found in the literature at finite temperature \cite{Varma-2002,Sondhi-1997}, in our case only the right-hand side of the fan can be identified. 
This is because on the left-hand side of the QCP, the FFLO critical line (full red line in Fig.~\ref{Figure-8}) remains stuck at $T=0$ (a possible finite-temperature transition line to a quasi-long-range ordered FFLO phase \cite{Radzihovsky-2011} being out of reach within the present approach). 
Note also that the non-Fermi-liquid region above $T^*_\text{NFL}$ (red shaded area in Fig.~\ref{Figure-8}) is limited to temperatures $T/T_F \lesssim 0.02$. 
This is because $T/T_F\simeq 0.02$ corresponds to a Matsubara frequency spacing $\Delta \omega_n =2\pi T \simeq 0.13$, which about corresponds to the upper limit below which the square-root non-Fermi-liquid behavior of the self-energies (as shown in Figs.~\ref{Figure-6} and \ref{Figure-7}) can be still resolved in a grid of discrete Matsubara frequencies.
It should further be remarked that the non-Fermi-liquid behavior induced by the proximity to a FFLO QCP in a density-imbalanced Fermi gas considered here is \emph{a priori\/} unrelated to the deviations from Fermi-liquid behavior that may even occur in a balanced system close to the superfluid critical temperature, due to the occurrence of a pseudo-gap in the normal phase \cite{Gaebler-2010,Perali-2011} (see also Ref.~\cite{Jensen-2019} for a recent review on the pseudo-gap in ultra-cold Fermi gases).

Having determined the crossover temperature $T^{*}_\text{NFL}$, we may now to look for deviations from Fermi-liquid theory at finite temperature by considering specifically the temperature behavior of the quasi-particle residues $Z_\sigma$.
To this end, at finite (albeit still rather low) temperature, the quasi-particle residues can be calculated in terms of the expression (cf., e.g., Refs.~\cite{Chen-2012,Liu-2018,Fontenele-2022})
\begin{equation}
Z_\sigma = \bigg[ 1- \frac{ \text{Im}  \Sigma_\sigma (k_{{\rm F} \sigma},i\omega_0)}{\omega_0}  \bigg]^{-1} \,,
\label{eq:residue_finiteT}
\end{equation}
where $\omega_0=\pi T$ is the first fermionic Matsubara frequency.  
For $T \ll T_{\rm F}$ Eq.~(\ref{eq:residue_finiteT}) well approximates an expression like (\ref{eq:residue}) (but now evaluated at finite $T$), namely,
 \begin{equation}
 Z_\sigma= \bigg[1- \frac{\partial \text{Im} \Sigma_\sigma(k_{{\rm F} \sigma}, i\omega)}{ \partial \omega} \bigg|_{\omega \to 0^{+}} \bigg]^{-1},
 \label{eq:residue_finiteT_analytic_continuation}
 \end{equation}
 which, however, would require an analytic continuation from the discrete Matsubara frequencies $i\omega_n$ to the continuous (imaginary) frequency $i \omega$.

In addition, the expression (\ref{eq:residue_finiteT}) can be used as a direct check of non-Fermi liquid behavior, by considering the expansion  
\begin{equation}
\text{Im} \Sigma^{(\text{FL})}_\sigma(k_{{\rm F} \sigma}, i\omega_n) \simeq (1-1/Z_\sigma) \omega_n + E_\sigma (\omega_n^2-\pi^2 T^2) 
\label{eq:ImSigma_expansion}
\end{equation}
where $E_\sigma $ is a constant, which is valid for small temperature and $\omega_n$.
For the first Matsubara frequency $\omega_0=\pi T$, this form implies that the quadratic term $\omega_0^2$ gets exactly compensated by the $\pi^2 T^2$ term (in accordance with the first Matsubara rule \cite{Chubukov-2012,Schafer-2021}).
For a Fermi liquid, the temperature dependence of $Z_\sigma(T) - Z_\sigma(T=0)$ as obtained from Eq.~(\ref{eq:residue_finiteT}) is expected to be super-linear (that is to say, of order higher than linear).
In particular, in 3D one expects $\mathcal{O}(T^2 \log T)$ \cite{Chubukov-2012}.
A different low-$T$ dependence of the expression (\ref{eq:residue_finiteT}) would then signal deviations from Fermi-liquid theory.

\begin{figure}[tb]  
\includegraphics[angle=0,width=1.0\columnwidth]{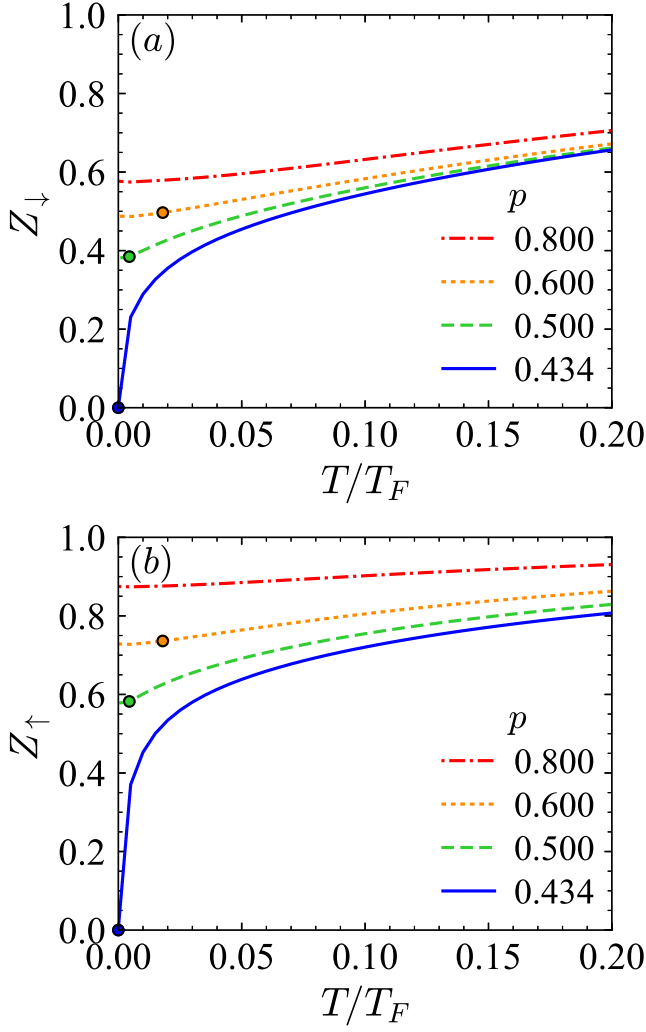}
\caption{Quasi-particle residues $Z_\sigma$ at unitarity as functions of temperature for both spin components $\sigma=(\uparrow,\downarrow)$ and several polarizations (where $p=p_c=0.434$ corresponds to the critical polarization 
             of the FFLO QCP at zero temperature). Circles indicate the points on the curves that correspond to the crossover temperatures $T^*_\text{NFL}$ obtained by Eq.~(\ref{eq:T*_NFL}) for given polarization.}
\label{Figure-9}
\end{figure}

Figure~\ref{Figure-9} shows the temperature dependence of the residues $Z_\sigma$ as obtained from Eq.~(\ref{eq:residue_finiteT}), at unitarity and for several polarizations. 
When $p = p_c=0.434$, the residues are seen to vanish for $T \to 0$ with a completely non-Fermi-liquid behavior. 
For polarizations $p>p_c$, the residues reach their finite $T=0$ value with a horizontal slope, corresponding to the super-linear behavior expected for a Fermi liquid.
However, it is interesting to note that, for polarizations $p=0.5$ and $0.6$ close enough to $p_c$, this super-linear behavior occurs only in a small temperature region close to $T=0$, after which a sub-linear behavior sets in. 
This region corresponds to temperatures $T \lesssim T^*_\text{NFL}$, where $T^*_\text{NFL}$ was independently identified by Eq.~(\ref{eq:T*_NFL}) for the same polarizations.
This shows the way how the crossover temperature $T^*_\text{NFL}$, that separates the non-Fermi-liquid from the Fermi-liquid regions, appears explicitly in a specific physical quantity. 

The temperature dependence of the residues $Z_\sigma$ at the critical polarization $p_c$ can be directly related to the dynamical scaling of the self-energies considered in Sec.~\ref{sec:Evolution_FL}.
If one assumes, like in Eq.~(\ref{eq:chi_pair_QCP_scalingT}), an $\tilde{\omega}/T$ scaling of the retarded self-energies at $p_c$ for low temperature, Eq.~(\ref{eq-sigmaR}) implies that
\begin{eqnarray}
\Delta \text{Re} \Sigma^{\text{R}}_\sigma(k_{{\rm F} \sigma},\tilde{\omega}; T) &\sim&T^{1/2} \phi_{1 \sigma} (\tilde{\omega}/T) 
\label{eq:ReSigma}
\\ 
\text{Im} \Sigma^{\text{R}}_\sigma(k_{{\rm F} \sigma},\tilde{\omega}; T)&\sim&T^{1/2} \phi_{2 \sigma} (\tilde{\omega}/T) 
\label{eq:SigmaR_Omega/T_scaling}
\end{eqnarray}
where $\phi_{1 \sigma} $ and $\phi_{2 \sigma}$ are universal scaling functions. 
The temperature dependence of the residues at $p_c$ and low $T$ is then obtained by using Eq.~(\ref{eq:ReSigma}) together with
\begin{equation}
Z_\sigma(T)= \left[1- \frac{\partial \textrm{Re} \Sigma^{\rm R}_\sigma(k_{{\rm F} \sigma}, \tilde{\omega}; T)}{\partial \tilde{\omega}} \bigg|_{\tilde{\omega} = 0} \right]^{-1} \, ,
\end{equation}
and reads
\begin{equation}
Z_\sigma(T) \simeq \frac{1}{1+G_\sigma T^{-1/2}}
\label{eq:Z_T}
\end{equation}
where $G_\sigma = |\phi_{1 \sigma}^{'}(0)|$ is a spin-dependent non-universal constant. 
By fitting the curves at $p=p_c=0.434$ with the form (\ref{eq:Z_T}), we obtain a robust fit with $R^2>0.9997$ (which extends up up to $T/T_F=0.1$).  
This result can be regarded as an indirect (albeit strong) check on the assumed $\tilde{\omega}/T$ scaling of Eqs.~(\ref{eq:ReSigma}) and (\ref{eq:SigmaR_Omega/T_scaling}).

Finally, we note that, by combining Eq.~(\ref{eq:SigmaR_Omega/T_scaling}) with Eq.~(\ref{eq:Z_T}) at the critical polarization $p_c$, one obtains an inverse quasi-particle lifetime at the Fermi surface $1/\tau_\sigma(k_{{\rm F} \sigma})$ 
scaling at low temperature like
\begin{equation}
\begin{split}
\frac{1}{\tau_\sigma(k_{{\rm F} \sigma})} \equiv& - Z_\sigma(T) \, \text{Im} \Sigma^{\text{R}}_\sigma (k_{{\rm F} \sigma}, \tilde{\omega}=0;T) \\
&\quad \sim \frac{T^{1/2}}{1+A_\sigma T^{-1/2}} \sim T \, ,
\end{split}
\end{equation}
which is linear rather than quadratic in $T$ as it should be in the Fermi liquid regime \cite{Negele-Orland-1988}. 
This non-Fermi liquid behavior of the quasi-particle lifetime is similar to what expected in condensed-matter systems in the proximity to a QCP, as produced either by charge \cite{Castellani-1995} or
magnetic \cite{Varma-2002,Lohneysen-2007} instabilities. 
In ultra-cold gases, this behavior could in principle be observed experimentally via radio-frequency spectroscopy \cite{Sagi-2015,Yan-2019}.


\section{Concluding remarks and outlook}
\label{sec:conclusions}
In this article, we have investigated the evolution of the normal phase of a two-component attractive Fermi gas in 3D at zero temperature, by varying the polarization from the polaronic limit to the superfluid QCP. 
Depending on coupling, this QCP is found to be either toward a homogeneous polarized phase (for strong coupling) or toward a inhomogeneous FFLO phase (for intermediate and weak couplings, including unitarity). 
In the case of a QCP toward a polarized BCS superfluid, we have found that a Fermi liquid description in terms of quasi-particles remains valid down to the QCP, while at an FFLO QCP we have found a complete breakdown of the quasi-particle description, similarly to what happens at an antiferromagnetic QCP in heavy-fermion materials. 

The non-Fermi liquid behavior at the FFLO QCP is characterized by the vanishing of the quasi-particle residues, the divergence of the quasi-particle effective masses, and an anomalous $\sim \omega^{1/2}$ frequency dependence of 
$\rm{Im} \Sigma$ and $\Delta \rm{Re} \Sigma$ at the Fermi surface.
This is consistent with a scaling behavior of the single-particle spectral function and of the dynamical pairing susceptibility at the QCP with a dynamical critical exponent $z=2$. 
At finite temperature, these results translate in a non-Fermi liquid critical region above the FFLO QCP, that presents an anomalous temperature dependence of physical quantities, like a sub-linear (rather than super-linear) temperature dependence of the quasi-particle residues and a linear (rather than quadratic) temperature dependence of the inverse quasi-particle lifetimes at the Fermi surface. 
The crossover temperature $T^*_\text{NFL}$, that corresponds to the boundary of this non-Fermi liquid region, has also been identified in the temperature-vs-polarization phase diagram at unitarity, by comparing the time-scales of thermal and quantum fluctuations. 

By combining our  predictions for the quantum phase transition between the normal and FFLO phases with the experimental and QMC data, we have determined that the FFLO QCP gets not covered by phase separation
in an extended coupling region from weak coupling up to $(k_{\rm F} a_{\rm F})^{-1} \simeq 0.1$, thereby including the unitary regime where $(k_{\rm F} a_{\rm F})^{-1}=0$. 
This finding implies that the FFLO QCP, with its non-Fermi liquid precursor features when coming from the normal phase, could (at least in principle) be directly accessed experimentally in a unitary Fermi gas.

This awaited experimental search for the FFLO QCP with ultra-cold Fermi gases would complement the alternative search going on in condensed matter.
In this respect, a recent article has reported convincing signatures of the FFLO phase in the material strontium ruthenate via nuclear magnetic resonance measurements \cite{FFLO-experimental-NMR}.
These measurements were performed at a temperature which is less than $5\%$ the value of the superfluid critical temperature of the sample, which can then be regarded essentially as zero temperature for all practical purposes.
Correspondingly, measurements with ultra-cold Fermi gases, where such low temperatures (relative to the critical temperature) are hard to reach even at unitarity (and especially for finite polarization), could take advantage
of the precursor effects arising in the normal phase that we have here predicted to occur even at finite (albeit still small) temperature.  In this respect, experiments performed with Fermi gases in box-like traps appear to be most promising \cite{Mukherjee-2017,Hueck-2018,Shkredov-2021,Navon-2021,Navon-2022}. 

In this context, it should be emphasized that our finding here for a polarized unitary Fermi gas, of a Fermi-liquid behavior at zero (or even low) temperature outside the classical fluctuation region around the FFLO QCP, does not contradict 
previous claims about the non-Fermi-liquid behavior of the same two-component but unpolarized Fermi gas at unitarity above its critical temperature \cite{Perali-2011}, 
whereby pseudo-gap effects and temperature broadening act (as expected) to wash up all fundamental features of a Fermi-liquid phase.
The two distinct regions of the temperature-vs-polarization phase diagram at unitarity, of interest here and in Ref.~\cite{Perali-2011}, are clearly apparent from Fig.~\ref{Figure-8} above.

Finally, it is worth emphasizing again that to obtain our results it was essential to employ a fully self-consistent $t$-matrix approach.
This is because this approach duly satisfies Luttinger theorem and thus correctly identifies the Fermi surface(s) about which the fermionic quasi-particles are defined.
In this context, it was also necessary to perform our numerical calculations \emph{exactly\/} at zero temperature.
This kind of technical requirement has led us to face and overcome a number of nontrivial challenges to render it computationally feasible; yet, it has proved essential to correctly identify a number of key physical features occurring in the evolution of the Fermi liquid with polarization.
This implementation exactly at zero temperature has never been attempted before, and represents the novel state-of-the-art computational technology in the field.

\begin{acknowledgments}
Partial financial support from the Italian MIUR under Project PRIN2017 (20172H2SC4) and next Generation EU and Italian MUR under project “National Centre for HPC, Big Data and Quantum Computing”, Contract: CN00000013 are acknowledged. M. P. is grateful to M. Zaccanti for support and discussions. 
\end{acknowledgments}

\appendix


\section{DETAILS OF THE NUMERICAL PROCEDURES}
\label{app:details_numerical_T=0}
In this Appendix, we first provide the details of the numerical procedures required to implement the cycles of self-consistency for a polarized attractive Fermi gas at zero temperature.  In the last section we then briefly describe the fitting procedure used to define the crossover polarization $p^*$ introduced in Fig.~\ref{Figure-5} of the main text.

In view of possible future extensions, when describing the numerical procedures to achieve self-consistency, we shall also consider the occurrence of a mass imbalance (although not included in this article) between the two components of the Fermi gas,
by introducing in the expressions below the factors $\gamma_\sigma=2m_{r}/m_{\sigma}$, where $m_\sigma$ is the mass of the $\sigma=(\uparrow,\downarrow)$ component and 
$m_{r}= m_\uparrow m_\downarrow/(m_\downarrow+m_\uparrow)$ is the associated reduced mass. 
In the mass-balanced case here considered, one has $2 m_{r}=m_\uparrow=m_\downarrow=m$ and  $\gamma_\sigma=1$. 

The procedures presented here, which rely on Fourier transforming back and forth between the $(\mathbf{k},i\omega)$ or $(\mathbf{Q},i\Omega)$ spaces and the $(\mathbf{r},\tau)$ space, are similar to those implemented at finite temperature in previous works \cite{Haussmann-1994,Haussmann-2007,FrankThesis-2018,PPS-2019}. 
Some important features, however, have to be modified to treat the zero-temperature limit correctly. 
For this reason, in the following we shall mostly focus on the specific details of this implementation at zero temperature, while the basic structure of the cycle of self-consistency can be found in Ref.~\cite{PPS-2019}.

All the expressions reported in this Appendix are given in dimensionless units, such that wave vectors are in units of the effective Fermi wave vector $k_{\rm F}=(3 \pi^2 n)^{1/3}$ (where $n=n_\uparrow+n_\downarrow$ is the total particle density) and energies are in units of the effective Fermi energy $E_{\rm F}=k_{\rm F}^{2}/(4m_{r})$. 
Accordingly, the fermionic single-particle propagators $G_\sigma(\mathbf{k},i\omega_{n})$ are in units of $E_{\rm F}^{-1}$, the fermionic self-energies $\Sigma_\sigma(\mathbf{k},i\omega_n)$ in units of $E_{\rm F}$, 
and the particle-particle propagator $\Gamma(\mathbf{Q},i\Omega_{\nu})$ in units of $(2m_{r} k_{\rm F})^{-1}$. 
Finally, to further shorten the notation, we adopt the symbol $v=(k_{\rm F} a_{\rm F})^{-1}$ for the dimensionless coupling.

\subsection{Imaginary time interval}
\label{subsec:tau_interval}
A few preliminary considerations are in order, about the interval to be utilized for the imaginary time variable $\tau$. 
At a finite temperature $T$, the periodicity (anti-periodicity) of bosonic (fermionic) single-particle propagators with period $\beta=1/T$ enables one to work typically in the interval $(0,\beta)$,
such that the corresponding interval at zero temperature would be $(0,+\infty)$.  
However, this straightforward choice turns out to be inconvenient in practice, because the singularities for $\tau \to \beta^-$ of some functions [typically, the self-energies $\Sigma_\sigma(\mathbf{r},\tau)$] would be shifted to the region 
$\tau \to +\infty$ where they are difficult to be dealt with numerically. 
By adopting instead the finite-temperature interval $(-\beta/2,+\beta/2)$, which becomes $(-\infty,+\infty)$ at zero temperature, all relevant singularities occur for $\tau \to 0^-$ or $\tau \to 0^+$ and can be accounted for more efficiently.

\subsection{Transforming from $G_\sigma(\mathbf{k},i\omega)$ to $G_\sigma(\mathbf{r},\tau)$}
\label{subsec:transforming_G}
The first functions to be Fourier transformed in the cycle of self-consistency are the single-particle Green's functions $G_\sigma$. 
Like for the finite-temperature case \cite{PPS-2019}, the Fourier transform can be performed in two steps, namely,
\begin{equation}
G_\sigma(\mathbf{k},i\omega) \stackrel{{\rm FT}}{\rightarrow} G_\sigma(\mathbf{k},\tau) \stackrel{{\rm FT}}{\rightarrow} G_\sigma(\mathbf{r},\tau) \, ,
\label{eq:G_FT_scheme}
\end{equation}
with the Fourier transform over the wave vector $\mathbf{k}$ following that over the frequency $\omega$. 
To perform these numerical Fourier transforms, it is helpful to add and subtract some functions known analytically in advance, in both $(\mathbf{k},i\omega)$ and $(\mathbf{r},\tau)$ representations. 
In the following, we will consider two different subtraction schemes: 
A primary one to deal with the large $(\mathbf{k},i\omega)$ behavior of $G_\sigma$, and a secondary one to deal with the quasi-particle contribution to $G_\sigma$ that accounts for the sharpness of the Fermi surface 
at $|\mathbf{k}|=k_{{\rm F} \sigma}$ and for the ensuing Friedel oscillations at large $\mathbf{r}$ in the $(\mathbf{r},\tau)$ representation.

\subsubsection{Primary subtraction scheme for \\ the large $(\mathbf{k},i\omega)$ behavior}
We begin by defining an analytic reference function $G_\sigma^{(a)}$ which embodies the leading large $\mathbf{k},i\omega$ (or, equivalently, the  small $|\mathbf{r}|$ and $\tau \to 0$) behavior of $G_\sigma$. 
The form of this analytic function is suggested by the free-particle propagator $G_{0 \sigma}$, and is given by
\begin{eqnarray}
G_\sigma^{(a)}(\mathbf{k},i\omega) &=& \frac{1}{i \omega -\gamma_\sigma \mathbf{k}^2 + \mu_{0\sigma}}  \nonumber \\
 &- & \frac{\Delta \mu_\sigma}{(i \omega - \gamma_\sigma \mathbf{k}^2 + \mu_{0\sigma})^2} 
 \label{eq:Ga(k,omega)} 
\end{eqnarray}
where the suffix $(a)$ stands for \emph{analytic}, with Fourier transforms
\begin{align}
G_\sigma^{(a)}(\mathbf{k},\tau) &= \int_{-\infty}^{+\infty} \frac{d \omega}{2\pi} G_\sigma^{(a)}(\mathbf{k},i\omega) e^{-i\omega \tau} \\
&= - (1+ \Delta \mu_\sigma \tau)  \, e^{-(\gamma_\sigma \mathbf{k}^2- \mu_{0\sigma}) \tau} \, \theta(\tau) \, , \label{eq:Ga(k,tau)} \\
G_\sigma^{(a)} (\mathbf{r},\tau) &= \int \frac{d \mathbf{k}}{(2\pi)^3} G_\sigma^{(a)}(\mathbf{k},\tau) e^{i \mathbf{k}\cdot \mathbf{r}} \\
&=- (1+ \Delta \mu_\sigma \tau) \frac{e^{\mu_{0\sigma} \tau} e^{-\frac{\mathbf{r}^2}{4\gamma_\sigma \tau}}}{(4\pi \gamma_\sigma \tau)^{3/2}} \, \theta(\tau) \, . 
\label{eq:Ga(r,tau)}
\end{align}
Here, $\theta(\tau)$ is the Heaviside step function, $\mu_{0 \sigma}$ is a negative auxiliary chemical potential for which we shall typically adopt the convenient expression $\mu_{0 \sigma} = -0.1-v^2 \theta(v)$, 
and $\Delta \mu_\sigma = \mu_\sigma - \mu_{0 \sigma}$ is the difference between the true and auxiliary chemical potentials. 
The introduction of the (negative) auxiliary chemical potential $\mu_{0 \sigma}$ avoids the divergence of $G^{(a)}_\sigma(\mathbf{k},i\omega)$ for $\omega \to 0$, and the second term in Eq.~(\ref{eq:Ga(k,omega)}) 
makes faster the decay to zero of the difference between $G_\sigma(\mathbf{k},i\omega)$ and $G_\sigma^{(a)}(\mathbf{k},i\omega)$ at large $\omega$ (thus leading to an $\omega^{-5/2}$ behavior). 

We then define the difference function 
\begin{equation}
G^{(n)}_\sigma(\mathbf{k}, i\omega) = G_\sigma(\mathbf{k}, i\omega) -G_\sigma^{(a)} (\mathbf{k}, i\omega) 
\label{eq:Gn(k,omega)}
\end{equation}
where the suffix $(n)$ stands for \emph{numeric}, and Fourier transform it first over $\omega$ and then over $\mathbf{k}$. 
Once the Fourier transforms over $\omega$ and $\mathbf{k}$ of Eq.~(\ref{eq:Gn(k,omega)}) are performed, $G_\sigma(\mathbf{r},\tau)$ is eventually obtained by adding $G_\sigma^{(a)}(\mathbf{r},\tau)$ given by Eq.~(\ref{eq:Ga(r,tau)}) 
to the transformed $G^{(n)}_\sigma(\mathbf{r},\tau)$.

\subsubsection{Secondary subtraction scheme for \\ the quasi-particle contribution}
Assuming that one also knows the quasi-particle residues $Z_\sigma$ and the effective masses $m^*_\sigma$ from a previous iteration of the loop of self-consistency (as calculated from the self-energy through Eqs.~(\ref{eq:residue}) and (\ref{eq:effective_mass})), a further refinement of the calculation can be made by analytically treating in Eq.~(\ref{eq:Gn(k,omega)}) also the quasi-particle contribution
\begin{equation}
G^{(\text{QP})}_\sigma(\mathbf{k},i\omega)= \frac{Z_\sigma}{i\omega - \gamma^*_\sigma (\mathbf{k}^2-k_{{\rm F} \sigma}^2) } \,,
\label{eq:Gqp_k_omega}
\end{equation}
which corresponds to a free quasi-particle propagator with residue $Z_\sigma$, effective mass $m^*_\sigma$ (such that $\gamma^*_\sigma=2m_{r}/m^*_{\sigma}$), 
and chemical potential $\mu_\sigma=\gamma^*_\sigma k_{{\rm F} \sigma}^2$ (note that, in general, $\gamma^*_\sigma \neq 1$ also in the mass-balanced case). 
This function allows us to analytically take into account the pole of $G_\sigma(\mathbf{k},i\omega)$ which is responsible for the discontinuity of $G_\sigma(\mathbf{k},\tau)$ at the Fermi surface $|\mathbf{k}|=k_{{\rm F} \sigma}$.  
[Note that the Fermi wave vector for the $\sigma$ component becomes $k_{\mathrm{F} \sigma}=(2n_\sigma/n)^{1/3}$ in dimensionless units.]

Note, however, that if one would merely subtract the quasi-particle contribution (\ref{eq:Gqp_k_omega}) in Eq.~(\ref{eq:Gn(k,omega)}), one would spoil the large-frequency \mbox{$\omega^{-5/2}$} behavior of the difference function $G^{(n)}_\sigma(\mathbf{k}, i\omega)$ obtained above with the primary subtraction scheme. 
It is thus necessary to subtract in Eq.~(\ref{eq:Gn(k,omega)}) a reference function that embodies the polar structure of Eq.~(\ref{eq:Gqp_k_omega}), but at the same time keeps an $\omega^{-5/2}$ tail for large frequencies. 
This reference function is then chosen as follows
\begin{equation}
G^{(\text{QP},n)}_\sigma(\mathbf{k},i\omega)=G^{(\text{QP})}_\sigma(\mathbf{k},i\omega)-G^{(\text{QP},a)}_\sigma(\mathbf{k},i\omega) \,,
\label{eq:Gqpn_k_omega}
\end{equation}
where, like in Eq.~(\ref{eq:Ga(k,omega)}), we define
\begin{eqnarray}
G^{(\text{QP},a)}_\sigma(\mathbf{k},i\omega) &=& \frac{Z_\sigma }{i \omega -\gamma^*_\sigma \mathbf{k}^2 + \mu_{0\sigma}}  \nonumber \\
 &-& \frac{Z_\sigma \Delta \mu^{(\text{QP})}_\sigma}{(i \omega - \gamma^*_\sigma \mathbf{k}^2 + \mu_{0\sigma})^2} 
\label{eq:Gqpa_k_omega} 
\end{eqnarray}
with $\Delta \mu^{(\text{QP})}_\sigma=\gamma^*_\sigma k_{F \sigma}^2-\mu_{0 \sigma}$ and $\mu_{0 \sigma}$ defined like in Eq.~(\ref{eq:Ga(k,omega)}). 
Similarly to the free Green's function, Eq.~(\ref{eq:Gqpn_k_omega}) can now be Fourier transformed analytically to the $(\mathbf{k},\tau)$ representation, to obtain
\begin{equation}
\begin{split}
G^{(\text{QP},n)}_\sigma(\mathbf{k},\tau) &= G^{(\text{QP},n,+)}_\sigma(\mathbf{k},\tau) \theta(\tau)\\
 &+ G^{(\text{QP},n,-)}_\sigma(\mathbf{k},\tau) \theta(-\tau) 
\end{split}
\label{eq:Gqpn_k_tau}
\end{equation}
where
\begin{equation}
\begin{split}
 &G^{(\text{QP},n,+)}(\mathbf{k},\tau) = Z_\sigma e^{-\gamma^*_\sigma (\mathbf{k}^2-k_{{\rm F} \sigma}^2) \tau} [\theta(k_{{\rm F} \sigma}-|\mathbf{k}|)-1] \\
& \quad  \quad + Z_\sigma  (1+ \Delta \mu^{(\text{QP})}_\sigma \tau)  \, e^{-(\gamma^*_\sigma \mathbf{k}^2- \mu_{0\sigma}) \tau},\\
 &G^{(\text{QP},n,-)}(\mathbf{k},\tau) = Z_\sigma e^{-\gamma^*_\sigma (\mathbf{k}^2-k_{{\rm F} \sigma}^2) \tau}\theta(k_{{\rm F} \sigma}-|\mathbf{k}|)  \, .
 \label{eq:Gqpn_k_tau_2}
\end{split}
\end{equation}

At this point the Fourier transform over $\mathbf{k}$ to get $G_\sigma^{(\text{QP},n)}(\mathbf{r},\tau)$ can be readily performed numerically (or, alternatively, one may utilize an analytic expression in terms of the complex error function).

Finally, one can define a refined difference function, in the form
\begin{equation}
G^{(n)}_\sigma(\mathbf{k}, i\omega) = G_\sigma(\mathbf{k}, i\omega) -G_\sigma^{(a)} (\mathbf{k}, i\omega)- G_\sigma^{(\text{QP}, n)}(\mathbf{k},i\omega) \, ,
\label{eq:Gn(k,omega)2}
\end{equation}
and Fourier transform it first over $\omega$ and then over $\mathbf{k}$. 
Once these Fourier transforms are performed, one can eventually obtain $G_\sigma(\mathbf{r},\tau)$ by adding back 
$G_\sigma^{(a)}(\mathbf{r},\tau)$ of Eq.~(\ref{eq:Ga(r,tau)}) and the just calculated $G_\sigma^{(\text{QP}, n)}(\mathbf{r},\tau)$ to the transformed $G^{(n)}_\sigma(\mathbf{r},\tau)$.

As far as the Fourier transform of the particle-particle bubble of Section~\ref{subsec:transforming_Rpp} below is concerned, it is once again useful to calculate the full $G_\sigma^{(\text{QP})}(\mathbf{r},\tau)$. 
This is obtained by adding back to $G_\sigma^{(\text{QP},n)}(\mathbf{r},\tau)$ the analytic contribution $G_\sigma^{(\text{QP},a)}$ in the $(\mathbf{r},\tau)$ representation, which in analogy with Eq.~(\ref{eq:Ga(r,tau)}) is given by:
\begin{equation}
\begin{split}
G_\sigma^{(\text{QP},a)} (\mathbf{r},\tau) =& - Z_\sigma (1+ \Delta \mu^{\text{(QP)}}_\sigma \tau) \\
&\times \frac{e^{\mu_{0\sigma} \tau} e^{-\frac{\mathbf{r}^2}{4\gamma^*_\sigma \tau}}}{(4\pi \gamma^*_\sigma \tau)^{3/2}} \theta(\tau) \, .
\end{split}
\label{eq:Gqpa_r_tau}
\end{equation}

\subsection{Transforming from $R_\mathrm{pp}(\mathbf{r},\tau)$ to $R_\mathrm{pp}(\mathbf{Q},i\Omega)$}
\label{subsec:transforming_Rpp}
An additional function to be Fourier transformed in the self-consistent cycle is the renormalized particle-particle bubble $R_\mathrm{pp}$ defined by Eq.~(\ref{eq:Rpp_Q}). 
Here, again, the Fourier transform is performed in two steps, namely,
\begin{equation}
R_\mathrm{pp}(\mathbf{r},\tau) \stackrel{{\rm FT}}{\rightarrow} R_\mathrm{pp}(\mathbf{Q},\tau) \stackrel{{\rm FT}}{\rightarrow}  R_\mathrm{pp}(\mathbf{Q},i\Omega) \, ,
\label{eq:Rpp_FT_scheme}
\end{equation}
with the Fourier transform over the imaginary time $\tau$ following that over the spatial position $\mathbf{r}$.

\subsubsection{Primary subtraction scheme for \\ the large $(\mathbf{Q},i\Omega)$ behavior}
In Eq.~(\ref{eq:Rpp_Q}) $R_\mathrm{pp}$ is expressed in terms of the single-particle Green's functions $G_\sigma$ in $(\mathbf{Q},i\Omega)$ space. 
If one Fourier transforms Eq.~(\ref{eq:Rpp_Q}) (in units of $4m_{r} k_\mathrm{F}$) to the $(\mathbf{r},\tau)$ space, the convolution therein becomes a simple product and one ends up with the expression:
\begin{equation}
R_\mathrm{pp}(\mathbf{r},\tau)= G_{\sigma} (\mathbf{r},\tau)  G_{\bar{\sigma}} (\mathbf{r},\tau)-\Lambda \delta(\mathbf{r}) \delta(\tau) \, .
\label{eq:Rpp_r_tau}
\end{equation}
Here, $\Lambda=\int_{|\mathbf{k}|<k_0} d \mathbf{k}/(2\pi)^3 1/(2\mathbf{k}^{2})$ is a constant that diverges with the ultraviolet cutoff $k_{0}$ of the integral over $\mathbf{k}$ in Eq.~(\ref{eq:Rpp_Q})  
(where $2m_{r}/\mathbf{k}^2$ becomes $1/(2\mathbf{k}^2)$ in dimensionless units). 
Also in the present case, the leading behavior for  $\tau\to 0$ and small $\mathbf{r}$ needs to be subtracted before performing the Fourier transform. 
To this end, we recall from Eq.~(\ref{eq:Ga(r,tau)}) that such a behavior is already known for $G_\sigma(\mathbf{r},\tau)$, so that it is convenient to define the function
\begin{equation}
\begin{split}
&R^{(a)}_\mathrm{pp} (\mathbf{r},\tau)= G_\sigma^{(a)} (\mathbf{r},\tau) G_{\bar{\sigma}}^{(a)} (\mathbf{r},\tau) -\Lambda \delta(\mathbf{r}) \delta(\tau) \\
&=E_0(\mathbf{r},\tau)+ \Delta \mu E_1 (\mathbf{r},\tau) + [(\Delta \mu)^2-(\Delta h)^2] E_2(\mathbf{r},\tau) \, .
\end{split}
 \label{eq:Rppa_r_tau}
\end{equation}
In this expression, $\Delta \mu= \mu - \mu_0$ and $\Delta h= h - h_0$ where $\mu= (\mu_\uparrow+\mu_\downarrow)/2$, $\mu_0= (\mu_{0\uparrow}+\mu_{0\downarrow})/2$, $h=(\mu_\uparrow-\mu_\downarrow)/2$, and $h_0=(\mu_{0\uparrow}-\mu_{0\downarrow})/2$. 
In addition, in the expression (\ref{eq:Rppa_r_tau}) we have also introduced the functions
\begin{align}
&E_0(\mathbf{r},\tau)=  \frac{ e^{2 \mu_0 \tau} e^{-\frac{\mathbf{r}^2}{2 \gamma_\uparrow \gamma_\downarrow \tau}}}{(\gamma_\uparrow \gamma_\downarrow )^{3/2} (4\pi)^3 \tau^3} \theta(\tau) -\Lambda \delta(\mathbf{r}) \delta(\tau) \, , \\
&E_1(\mathbf{r},\tau)=   2 \frac{ e^{2 \mu_0 \tau} e^{-\frac{\mathbf{r}^2}{2 \gamma_\uparrow \gamma_\downarrow \tau}} }{(\gamma_\uparrow \gamma_\downarrow )^{3/2} (4\pi)^3 \tau^2} \theta(\tau)\, , \\
&E_2(\mathbf{r},\tau)=  \frac{ e^{2 \mu_0 \tau} e^{-\frac{\mathbf{r}^2}{2 \gamma_\uparrow \gamma_\downarrow \tau}}}{(\gamma_\uparrow \gamma_\downarrow )^{3/2} (4\pi)^3 \tau} \theta(\tau) \, .
\end{align}
By Fourier transforming these expressions to the $(\mathbf{Q},i\Omega)$ space, we obtain
\begin{align}
\label{eq:E0_Q_Omega}
&E_0(\mathbf{Q},i\Omega)= - \frac{1}{8 \sqrt{2} \pi} (\gamma_\uparrow \gamma_\downarrow \mathbf{Q}^2/2-2 \mu_0 - i \Omega)^{1/2} \, , \\
\label{eq:E1_Q_Omega}
&E_1(\mathbf{Q},i\Omega)= \frac{1}{8 \sqrt{2} \pi}  (\gamma_\uparrow \gamma_\downarrow \mathbf{Q}^2/2-2 \mu_0 - i \Omega)^{-1/2}\, , \\
\label{eq:E2_Q_Omega}
&E_2(\mathbf{Q},i\Omega)= \frac{1}{32 \sqrt{2} \pi}  (\gamma_\uparrow \gamma_\downarrow \mathbf{Q}^2/2-2 \mu_0 - i \Omega)^{-3/2} ,
\end{align}
so that $R^{(a)}_\mathrm{pp}$ of Eq.~(\ref{eq:Rppa_r_tau}) reads in $(\mathbf{Q},i\Omega)$ space:
\begin{equation}
\begin{split}
R^{(a)}_\mathrm{pp}&(\mathbf{Q},i\Omega) = E_0(\mathbf{Q},i\Omega) \\
& + \Delta \mu E_1(\mathbf{Q},i\Omega) + [ (\Delta \mu )^2-(\Delta h)^2 ] E_2(\mathbf{Q},i\Omega) \, .
\end{split}
\label{eq:Rppa_Q_Omega}
\end{equation}
Note that the divergent constant $\Lambda$ of Eq.~(\ref{eq:Rppa_r_tau}) has been reabsorbed in the definition of $E_0$. 
This is because in $(\mathbf{Q},i\Omega)$ space the regularizing term $-2m_{r}/\mathbf{k}^2$ of Eq.~(\ref{eq:Rpp_Q}) (which reads $-1/(2\mathbf{k}^2)$ in dimensionless units) is duly taken into account in $E_0$ since
\begin{equation}
\begin{split}
&E_0 (\mathbf{Q},i\Omega) = \int \!\!\!\frac{d\mathbf{k}}{(2\pi)^3}\bigg\{  \int \frac{d \omega}{2 \pi}  \bigg[ G^{(a)}_\sigma(\mathbf{k},i\omega)|_{\mu_\sigma=\mu_{0\sigma}}   \\
&\quad \times G^{(a)}_{\bar{\sigma}} (\mathbf{Q}-\mathbf{k},i\Omega-i\omega)|_{ \mu_{\bar{\sigma}}=\mu_{0\bar{\sigma}}}\bigg]- \frac{1}{2 \mathbf{k}^2}\bigg\} 
\end{split}
\end{equation}
where $G^{(a)}_\sigma(\mathbf{k},i\omega)|_{\mu_\sigma=\mu_{0\sigma}} =(i \omega - \gamma_\sigma \mathbf{k}^2 + \mu_{0 \sigma})^{-1}$ is the first term of Eq.~(\ref{eq:Ga(k,omega)}).

Now that the leading contribution of Eq.~(\ref{eq:Rpp_r_tau}) for $\tau\to 0$ and small $\mathbf{r}$ (as determined by the quadratic term $G_\sigma^{(a)}G_{\bar{\sigma}}^{(a)}$) has been taken into account, 
we can also take into account the sub-leading contribution as determined by the linear terms in $G_\sigma^{(a)}$. 
By expressing the Green's functions $G_\sigma$ in Eq.~(\ref{eq:Rpp_r_tau}) as $ G^{(n)}_\sigma + G_\sigma^{(a)}$ like in Eq.~(\ref{eq:Gn(k,omega)}) and retaining only the linear terms in $G_\sigma^{(a)}$, we are led to identify the following contribution
\begin{equation}
R_\mathrm{pp}^{(sa)}(\mathbf{r},\tau) = \sum_\sigma G^{(n)}_{\bar{\sigma}}(\mathbf{r}=0,\tau) \, G_\sigma^{(a)}(\mathbf{r},\tau) \, ,
\label{eq:Rppsa_Q_Omega}
\end{equation}
where we have set $\mathbf{r}=0$ in $G^{(n)}_{\bar{\sigma}}$ because from Eq.~(\ref{eq:Ga(r,tau)}) $G_\sigma^{(a)}$ is peaked about $\mathbf{r}=0$ for $\tau \to 0^+$ (while $G^{(n)}_{\bar{\sigma}}$ is a smooth function of $\mathbf{r}$). 
The suffix $(sa)$ here stands for \emph{semi-analytic}, because although $G_\sigma^{(a)}$ is analytic its coefficient $G^{(n)}_\sigma$ is determined numerically. 
The expression (\ref{eq:Rppsa_Q_Omega}) can be transformed to the $(\mathbf{Q},\tau)$ space, to obtain
\begin{equation}
R_\mathrm{pp}^{(sa)}(\mathbf{Q},\tau) = \sum_\sigma G^{(n)}_{\bar{\sigma}}(\mathbf{r}=0,\tau) \, G_\sigma^{(a)}(\mathbf{Q},\tau) \, ,
\label{eq:Rppsa_Q_tau}
\end{equation}
where $G_\sigma^{(a)}(\mathbf{Q},\tau)$ is given by Eq.~(\ref{eq:Ga(k,tau)}) with  $\mathbf{Q}$ replacing $\mathbf{k}$.
We can finally define the difference function
\begin{equation}
\Delta R_\mathrm{pp} (\mathbf{r},\tau) = R_\mathrm{pp} (\mathbf{r},\tau) - R_\mathrm{pp}^{(a)} (\mathbf{r},\tau) - R_\mathrm{pp}^{(sa)} (\mathbf{r},\tau) 
\label{eq:DRpp_r_tau}
\end{equation}
and Fourier transform it over $\mathbf{r}$ to obtain $\Delta R_\mathrm{pp} (\mathbf{Q},\tau)$.

As far as the Fourier transform from $\tau$ to $\Omega$ is concerned, we note that the semi-analytic term of Eq.~(\ref{eq:Rppsa_Q_Omega}) cannot be easily transformed. 
It is thus necessary to isolate its leading behavior for $\tau \to 0^+$ given by
\begin{equation}
\tilde{R}_\mathrm{pp}^{(sa)}(\mathbf{Q},\tau)  =  \sum_\sigma n_{\bar{\sigma}} \, G_\sigma^{(a)}(\mathbf{Q},\tau) \, .
\label{eq:Rppsa'_Q_tau}
\end{equation}
This expression is obtained by setting $\tau=0^+$ in the coefficient $G^{(n)}_{\bar{\sigma}}(\mathbf{r}=0,\tau)$ of Eq.~(\ref{eq:Rppsa_Q_tau}) and then identifying it with the density $n_{\bar{\sigma}}$ of the $\bar{\sigma}$ component. 
The Fourier transform of Eq.~(\ref{eq:Rppsa'_Q_tau}) over $\tau$ is thus given by
\begin{equation}
\tilde{R}_\mathrm{pp}^{(sa)}(\mathbf{Q},i\Omega)  =  \sum_\sigma n_{\bar{\sigma}} \, G_\sigma^{(a)}(\mathbf{Q},i\Omega) \, ,
\label{eq:Rppsa'_Q_Omega}
\end{equation}
where $G_\sigma^{(a)}(\mathbf{Q},i\Omega)$ is obtained from the expression (\ref{eq:Ga(k,omega)}) by replacing $\mathbf{k}$ and $\omega$ by $\mathbf{Q}$ and $\Omega$, respectively. 
Next we define a new difference function
\begin{equation}
\begin{split}
\tilde{\Delta}& R_\mathrm{pp}(\mathbf{Q},\tau)=  \Delta R_\mathrm{pp}(\mathbf{Q},\tau) \\
&\quad + R_\mathrm{pp}^{(sa)}(\mathbf{Q},\tau) - \tilde{R}_\mathrm{pp}^{(sa)}(\mathbf{Q},\tau) \, ,
\end{split}
\label{eq:tildeDRpp_Q_tau}
\end{equation}
which can be Fourier transformed over $\tau$ numerically.
The quantity $R_\mathrm{pp} (\mathbf{Q},i\Omega)$ is eventually obtained (in line with Eq.~(\ref{eq:DRpp_r_tau})) by adding $R_\mathrm{pp}^{(a)} (\mathbf{Q},i\Omega)$ of Eq.~(\ref{eq:Rppa_Q_Omega}) and $\tilde{R}_\mathrm{pp}^{(sa)} (\mathbf{Q},i\Omega)$ of Eq.~(\ref{eq:Rppsa'_Q_Omega}) to $\tilde{\Delta} R_\mathrm{pp} (\mathbf{Q},i\Omega)$.

\subsubsection{Secondary subtraction scheme for \\ the quasi-particle contributions}
Similarly to the Fourier transform of $G_\sigma$, if the quasi-particle residues $Z_\sigma$ and effective masses $m^*_\sigma$ are known from a previous iteration of the self-consistency loop, one can refine the calculation by subtracting the quasi-particle contributions to $R_\mathrm{pp}$. 
This refinement is particularly useful in the present context, because it can partially eliminate the Friedel oscillations in the $\mathbf{r}$-dependence of Eq.~(\ref{eq:DRpp_r_tau}). 
These oscillations, which originate from the presence of sharp Fermi surfaces in the Green's functions $G_\sigma(\mathbf{k},\tau)$, are especially difficult to be integrated over because for large $\mathbf{r}$ their amplitude decays only polynomially.
The strategy here is thus to start from the quasi-particle Green's functions (\ref{eq:Gqp_k_omega}) and build the quasi-particle contribution to the renormalized particle-particle bubble as follows
\begin{equation}
\begin{split}
&R^{(\mathrm{QP})}_{\mathrm{pp}}(\mathbf{Q}, i\Omega) =  \int \!\!\!\frac{d\mathbf{k}}{(2\pi)^3} \bigg[ \int \frac{d \omega}{2 \pi} G^{(\mathrm{QP})}_\sigma(\mathbf{k},i\omega)  \\
&\quad \times G^{(\mathrm{QP})}_{\bar{\sigma}} (\mathbf{Q}-\mathbf{k},i\Omega-i\omega) -  \frac{Z_\uparrow Z_\downarrow}{(\gamma^*_\uparrow + \gamma^*_\downarrow)\mathbf{k}^2} \! \bigg] \,,
\label{eq:Rppqp_Q_omega}
\end{split}
\end{equation}
where recall that $\gamma^*_\sigma=2m_{r}/m^*_{\sigma}$.
This expression can be evaluated analytically (cf. also Eq.~(16) of Ref.~\cite{Fratini-2012}), and takes the form
\begin{equation}
\begin{split}
&R^{(\mathrm{QP})}_{\mathrm{pp}}(\mathbf{Q}, i\Omega) = R^{(\mathrm{QP},\mathrm{sc})}_{\mathrm{pp}}(\mathbf{Q},i\Omega) \\
&\quad + \Delta R^{(\mathrm{QP},\uparrow)}_{\mathrm{pp}}(\mathbf{Q}, i\Omega) + \Delta R^{(\mathrm{QP},\downarrow)}_{\mathrm{pp}}(\mathbf{Q},i \Omega) \, .
\label{eq:Rppqp_Q_Omega_terms}
\end{split}
\end{equation}
Here,
\begin{equation}
\begin{split}
&R^{(\mathrm{QP},\mathrm{sc})}_{\mathrm{pp}}(\mathbf{Q},i\Omega)= - \frac{Z_\uparrow Z_\downarrow}{4 (\gamma^*_\uparrow+ \gamma^*_\downarrow)^{3/2} \pi} \\
&\phantom{aaaaaaa}\times \bigg(\frac{\gamma^*_\uparrow \gamma^*_\downarrow}{\gamma^*_\uparrow+\gamma^*_\downarrow}\mathbf{Q}^2-2 \mu^{(\mathrm{QP})}-i \Omega \bigg)^{1/2}
\end{split}
\label{eq:Rppqpsc_Q_Omega}
\end{equation}
is the strong-coupling contribution which, similarly to Eq.~(\ref{eq:E0_Q_Omega}), represents the leading $(\mathbf{Q},i\Omega)\to + \infty$ behavior, and the two remaining terms are given by
\begin{equation}
\begin{split}
&\Delta R^{(\mathrm{QP},\sigma)}_{\mathrm{pp}}(\mathbf{Q}, i\Omega) = \frac{Z_\uparrow Z_\downarrow}{4\pi^2 (\gamma^*_\uparrow+\gamma^*_\downarrow)}  \bigg\{ k_{\mathrm{F} \sigma} \\
&+\frac{k_\Omega(\mathbf{Q})}{2 (\gamma^*_\uparrow+\gamma^*_\downarrow)^{1/2}}  \log \bigg[ \frac{k_\Omega(\mathbf{Q})-(\gamma^*_\uparrow+\gamma^*_\downarrow)^{1/2}k_{1 \sigma} (\mathbf{Q})}{k_\Omega(\mathbf{Q})+(\gamma^*_\uparrow+\gamma^*_\downarrow)^{1/2} k_{1 \sigma} (\mathbf{Q})} \\
& \times \frac{k_\Omega(\mathbf{Q})+(\gamma^*_\uparrow+\gamma^*_\downarrow)^{1/2}k_{2 \sigma} (\mathbf{Q})}{k_\Omega(\mathbf{Q})-(\gamma^*_\uparrow+\gamma^*_\downarrow)^{1/2} k_{2 \sigma} (\mathbf{Q})} \bigg] \bigg\} \, ,
\end{split}
\end{equation}

\noindent
with the short-hand notation
\begin{equation}
\begin{split}
&k_{1\sigma}(\mathbf{Q})= k_{{\rm F} \sigma}-\frac{\gamma^*_{\bar{\sigma}}}{\gamma^*_\uparrow+\gamma^*_\downarrow} |\mathbf{Q}| \, , \\
&k_{2\sigma}(\mathbf{Q})= k_{{\rm F} \sigma}+\frac{\gamma^*_{\bar{\sigma}}}{\gamma^*_\uparrow+\gamma^*_\downarrow} |\mathbf{Q}| \, , \\
&k_\Omega (\mathbf{Q}) = i \bigg(\frac{\gamma^*_\uparrow \gamma^*_\downarrow}{\gamma^*_\uparrow + \gamma^*_\downarrow} \mathbf{Q}^2 - 2 \mu^{\mathrm{(QP)}}-i \Omega \bigg)^{1/2} 
\end{split}
\end{equation}
and the definition $\mu^{(\mathrm{QP})}= (\gamma^*_\uparrow k_{{\rm F} \uparrow}^2+ \gamma^*_\downarrow k_{{\rm F} \downarrow}^2)/2$. 

In the representation $(\mathbf{r},\tau)$, Eq.~(\ref{eq:Rppqp_Q_omega}) becomes (cf.~Eq.~(\ref{eq:Rpp_r_tau}))
\begin{equation}
\begin{split}
R^{(\mathrm{QP})}_{\mathrm{pp}}(\mathbf{r},\tau) &= G^{(\mathrm{QP})}_{\sigma} (\mathbf{r},\tau)  G^{(\mathrm{QP})}_{\bar{\sigma}} (\mathbf{r},\tau) \\
&\quad -  \Lambda^{(\mathrm{QP})} \delta(\mathbf{r}) \delta(\tau) \, ,
\end{split}
\label{eq:Rppqp_r_tau}
\end{equation}
where the quasi-particle Green's functions $G^{(\mathrm{QP})}_{\sigma} (\mathbf{r},\tau)$ are obtained numerically by Fourier transforming Eq.~(\ref{eq:Gqpn_k_tau}) and adding back the analytic contribution (\ref{eq:Gqpa_r_tau}), 
while $\Lambda^{(\mathrm{QP})} = 2 \Lambda Z_\uparrow Z_\downarrow/(\gamma^*_\uparrow + \gamma^*_\downarrow)$ with the constant $\Lambda$ defined in Eq.~(\ref{eq:Rpp_r_tau}).

Note, however, that one cannot simply subtract the quasi-particle contribution (\ref{eq:Rppqp_r_tau}) in Eq.~(\ref{eq:DRpp_r_tau}), since one would otherwise reintroduce a singular behavior for $\tau \to 0^+$ and small $\mathbf{r}$. 
A solution to this problem is to define a non-singular auxiliary function obtained by removing from Eq.~(\ref{eq:Rppqp_r_tau}) the singular $\tau \to 0^+$ and small-$\mathbf{r}$ behavior (thereby adopting a procedure similar to that 
described above for $R_{\rm pp}^{(a)}$), and then use this function to subtract away the oscillating large-$\mathbf{r}$ behavior in Eq.~(\ref{eq:DRpp_r_tau}) without introducing any singularity for $\tau \to 0^+$ and small $\mathbf{r}$. 
Similarly to what we did in Eq.~(\ref{eq:Rppa_r_tau}), we then define the function
\begin{equation}
\begin{split}
&R^{(\mathrm{QP},a)}_\mathrm{pp} (\mathbf{r},\tau)=E^{(\mathrm{QP})}_0(\mathbf{r},\tau)+ \Delta \mu^{(\mathrm{QP})} E^{(\mathrm{QP})}_1 (\mathbf{r},\tau) \\
&+ \Big[ \big(\Delta \mu^{(\mathrm{QP})}\big)^2-\big(\Delta h^{(\mathrm{QP})}\big)^2 \Big] E^{(\mathrm{QP})}_2(\mathbf{r},\tau) ,
 \end{split}
 \label{eq:Rppqpa_r_tau}
\end{equation}
where now $\Delta \mu^{(\mathrm{QP})}= \mu^{(\mathrm{QP})} - \mu_0$ and $\Delta h= h^{(\mathrm{QP})} - h_0$, with $\mu^{(\mathrm{QP})}= (\gamma^*_\uparrow k_{{\rm F} \uparrow}^2+ \gamma^*_\downarrow k_{{\rm F} \downarrow}^2)/2$, $\mu_0= (\mu_{0\uparrow}+\mu_{0\downarrow})/2$, $h^{(\mathrm{QP})}= (\gamma^*_\uparrow k_{{\rm F} \uparrow}^2- \gamma^*_\downarrow k_{{\rm F} \downarrow}^2)/2$, and $h_0=(\mu_{0\uparrow}-\mu_{0\downarrow})/2$. 
In addition, in the expression (\ref{eq:Rppqpa_r_tau}) we have introduced the analytic functions
\begin{align}
E^{(\mathrm{QP})}_0(\mathbf{r},\tau) &= Z_\uparrow Z_\downarrow  \frac{ e^{2 \mu_0 \tau} e^{-\frac{(\gamma^*_\uparrow+\gamma^*_\downarrow)\mathbf{r}^2}{4 \gamma^*_\uparrow \gamma^*_\downarrow \tau}}}{(\gamma^*_\uparrow \gamma^*_\downarrow )^{3/2} (4\pi)^3 \tau^3} \theta(\tau) \nonumber \\
& \quad -\Lambda^{(\mathrm{QP})} \delta(\mathbf{r}) \delta(\tau) \, , \\
E^{(\mathrm{QP})}_1(\mathbf{r},\tau) &=   2 Z_\uparrow Z_\downarrow \frac{ e^{2 \mu_0 \tau} e^{-\frac{(\gamma^*_\uparrow+\gamma^*_\downarrow)\mathbf{r}^2}{4 \gamma^*_\uparrow \gamma^*_\downarrow \tau}} }{(\gamma^*_\uparrow \gamma^*_\downarrow )^{3/2} (4\pi)^3 \tau^2} \theta(\tau)\, , \\
E^{(\mathrm{QP})}_2(\mathbf{r},\tau) &= Z_\uparrow Z_\downarrow \frac{ e^{2 \mu_0 \tau} e^{-\frac{(\gamma^*_\uparrow+\gamma^*_\downarrow)\mathbf{r}^2}{4 \gamma^*_\uparrow \gamma^*_\downarrow \tau}}}{(\gamma^*_\uparrow \gamma^*_\downarrow )^{3/2} (4\pi)^3 \tau} \theta(\tau) \, .
\end{align}
By Fourier transforming these expressions to the $(\mathbf{Q},i\Omega)$ space, we obtain correspondingly:
\begin{align}
\label{eq:E0qp_Q_Omega}
E^{(\mathrm{QP})}_0(\mathbf{Q},i\Omega) &= - \frac{Z_\uparrow Z_\downarrow}{4 (\gamma^*_\uparrow+ \gamma^*_\downarrow)^{3/2} \pi} \nonumber \\
&\times \bigg(\frac{\gamma^*_\uparrow \gamma^*_\downarrow}{\gamma^*_\uparrow+\gamma^*_\downarrow}\mathbf{Q}^2-2 \mu_0-i \Omega \bigg)^{1/2} \, , \\
\label{eq:E1qp_Q_Omega}
E^{(\mathrm{QP})}_1(\mathbf{Q},i\Omega)&=  \frac{Z_\uparrow Z_\downarrow}{4 (\gamma^*_\uparrow+ \gamma^*_\downarrow)^{3/2} \pi} \nonumber \\
&\times \bigg(\frac{\gamma^*_\uparrow \gamma^*_\downarrow}{\gamma^*_\uparrow+\gamma^*_\downarrow}\mathbf{Q}^2-2 \mu_0-i \Omega \bigg)^{-1/2}\, , \\
\label{eq:E2qp_Q_Omega}
E^{(\mathrm{QP})}_2(\mathbf{Q},i\Omega)&=  \frac{Z_\uparrow Z_\downarrow}{16(\gamma^*_\uparrow+ \gamma^*_\downarrow)^{3/2} \pi} \nonumber \\
&\times \bigg(\frac{\gamma^*_\uparrow \gamma^*_\downarrow}{\gamma^*_\uparrow+\gamma^*_\downarrow}\mathbf{Q}^2-2 \mu_0-i \Omega \bigg)^{-3/2} \, .
\end{align}
In this way, the function $R^{(\mathrm{QP},a)}_\mathrm{pp}$ defined by Eq.~(\ref{eq:Rppqpa_r_tau}) becomes in $(\mathbf{Q},i\Omega)$ space: 
\begin{equation}
\begin{split}
R^{(\mathrm{QP},a)}_\mathrm{pp}&(\mathbf{Q},i\Omega) = E^{(\mathrm{QP})}_0(\mathbf{Q},i\Omega) +\Delta \mu^{(\mathrm{QP})} E^{(\mathrm{QP})}_1(\mathbf{Q},i\Omega) \\
& + \Big[ \big(\Delta \mu^{(\mathrm{QP})} \big)^2-\big(\Delta h^{(\mathrm{QP})})^2 \Big] E^{(\mathrm{QP})}_2(\mathbf{Q},i\Omega) \, .
\end{split}
\label{eq:Rppqpa_Q_Omega}
\end{equation}

In analogy to Eq.~(\ref{eq:Rppsa_Q_Omega}), one can further take into account the sub-leading behavior for $\tau \to 0^+$ and small $\mathbf{r}$, by considering the semi-analytic contribution
\begin{equation}
\begin{split}
R^{(\mathrm{QP},sa)}_\mathrm{pp}(\mathbf{r},\tau) &= \sum_\sigma G^{(\mathrm{QP},n)}_{\bar{\sigma}}(\mathbf{r}=0,\tau) \, G_\sigma^{(\mathrm{QP},a)}(\mathbf{r},\tau) \, ,
\label{eq:Rppqpsa_Q_Omega}
\end{split}
\end{equation}
where $G^{(\mathrm{QP},n)}_{\sigma}(\mathbf{r},\tau)=G^{(\mathrm{QP})}_{\sigma}(\mathbf{r},\tau)-G^{(\mathrm{QP},a)}_{\sigma}(\mathbf{r},\tau)$ and $G_\sigma^{(\text{QP},a)} (\mathbf{r},\tau)$ is given by Eq.~(\ref{eq:Gqpa_r_tau}).
The expression (\ref{eq:Rppqpsa_Q_Omega}) can be transformed to the $(\mathbf{Q},\tau)$ space, yielding
\begin{equation}
R_\mathrm{pp}^{(\mathrm{QP},sa)}(\mathbf{Q},\tau) = \sum_\sigma G^{(\mathrm{QP},n)}_{\bar{\sigma}}(\mathbf{r}=0,\tau) \, G_\sigma^{(\mathrm{QP},a)}(\mathbf{Q},\tau) \, ,
\label{eq:Rppqpsa_Q_tau}
\end{equation}
where in analogy to Eq.~(\ref{eq:Ga(k,tau)}) (cf.~also Eq.~(\ref{eq:Gqpn_k_tau_2}))
\begin{equation}
\begin{split}
G_\sigma^{(\mathrm{QP},a)}(\mathbf{Q},\tau) &= - Z_\sigma (1+ \Delta \mu^{(\mathrm{QP})}_\sigma \tau) \\
& \quad \times e^{-(\gamma^*_\sigma \mathbf{Q}^2- \mu_{0\sigma}) \tau} \, \theta(\tau) 
\label{eq:Gqpa(Q,tau)}
\end{split}
\end{equation}
with $\Delta \mu^{(\mathrm{QP})}_\sigma = \gamma^*_\sigma k_{F\sigma}^2 - \mu_{0 \sigma}$. 
One can thus define the refined difference function
\begin{equation}
\begin{split}
&\Delta R_\mathrm{pp} (\mathbf{r},\tau) = R_\mathrm{pp} (\mathbf{r},\tau) - R_\mathrm{pp}^{(a)} (\mathbf{r},\tau) - R_\mathrm{pp}^{(sa)} (\mathbf{r},\tau) \\
&-R^{(\mathrm{QP})}_\mathrm{pp} (\mathbf{r},\tau) +R^{(\mathrm{QP},a)}_\mathrm{pp} (\mathbf{r},\tau) + R^{(\mathrm{QP},sa)}_\mathrm{pp} (\mathbf{r},\tau) \, ,
\end{split}
\label{eq:DRpp_r_tau2}
\end{equation}
and Fourier transform it over $\mathbf{r}$ to obtain $\Delta R_\mathrm{pp} (\mathbf{Q},\tau)$. 

As far as the Fourier transform from $\tau$ to $\Omega$ is concerned, in analogy to Eq.~(\ref{eq:Rppsa'_Q_tau}) it is useful to define a function which accounts for the leading $\tau \to 0^+$ behavior of Eq.~(\ref{eq:Rppqpsa_Q_Omega}),
by writing
\begin{equation}
\tilde{R}_\mathrm{pp}^{(\mathrm{QP},sa)}(\mathbf{Q},\tau)  =  \sum_\sigma  n^{(\mathrm{QP})}_{\bar{\sigma}}  \, G_\sigma^{(\mathrm{QP},a)}(\mathbf{Q},\tau) \, .
\label{eq:Rppqpsa'_Q_tau}
\end{equation}
where $n^{(\mathrm{QP})}_{\bar{\sigma}}= G_{\bar{\sigma}}^{(\mathrm{QP},n)}(\mathbf{r}=0,\tau=0^+)$ by our definition.
This expression can be transformed to the $(\mathbf{Q},i\Omega)$ space, yielding
\begin{equation}
\tilde{R}_\mathrm{pp}^{(\mathrm{QP},sa)}(\mathbf{Q},i\Omega)  =  \sum_\sigma  n^{(\mathrm{QP})}_{\bar{\sigma}}  \, G_\sigma^{(\mathrm{QP},a)}(\mathbf{Q},i\Omega)
\label{eq:Rppqpsa'_Q_Omega}
\end{equation}
where $G_\sigma^{(\mathrm{QP},a)}(\mathbf{Q},i\Omega)$ is given by Eq.~(\ref{eq:Gqpa_k_omega}) with $\mathbf{Q}$ and $\Omega$ replacing $\mathbf{k}$ and $\omega$, respectively. 
In analogy to Eq.~(\ref{eq:tildeDRpp_Q_tau}), we can now define a new difference function in $(\mathbf{Q},\tau)$ space as 
\begin{equation}
\begin{split}
\tilde{\Delta}& R_\mathrm{pp}(\mathbf{Q},\tau)= \Delta R_\mathrm{pp}(\mathbf{Q},\tau) + R_\mathrm{pp}^{(sa)}(\mathbf{Q},\tau) - \tilde{R}_\mathrm{pp}^{(sa)}(\mathbf{Q},\tau)   \\
&-  R_\mathrm{pp}^{(\mathrm{QP},sa)}(\mathbf{Q},\tau) + \tilde{R}_\mathrm{pp}^{(\mathrm{QP},sa)}(\mathbf{Q},\tau)  \, ,
\end{split}
\label{eq:tildeDRppqp_Q_tau}
\end{equation}
Fourier transform it over $\tau$ to obtain $\tilde{\Delta} R_\mathrm{pp}(\mathbf{Q},i\Omega)$, and finally obtain $R_\mathrm{pp} (\mathbf{Q},i\Omega)$ as
\begin{equation}
\begin{split}
&R_\mathrm{pp} (\mathbf{Q},i\Omega) =\tilde{\Delta} R_\mathrm{pp}(\mathbf{Q},i\Omega)+R_\mathrm{pp}^{(a)} (\mathbf{Q},i\Omega)+ \tilde{R}_\mathrm{pp}^{(sa)} (\mathbf{Q},i\Omega)  \\
& \ + R_\mathrm{pp}^{(\mathrm{QP})} (\mathbf{Q},i\Omega)- R_\mathrm{pp}^{(\mathrm{QP},a)} (\mathbf{Q},i\Omega) - \tilde{R}_\mathrm{pp}^{(\mathrm{QP},sa)} (\mathbf{Q},i\Omega).
\end{split}
\end{equation}

\subsection{Transforming from $\Gamma(\mathbf{Q},i\Omega)$ to $\Gamma(\mathbf{r},\tau)$}
\label{subsec:transforming_Gamma}
The next function to be Fourier transformed in the cycle of self-consistency is the particle-particle propagator $\Gamma$. 
Also in this case, the Fourier transform is performed in two steps, namely,
\begin{equation}
\Gamma(\mathbf{Q},i\Omega) \stackrel{{\rm FT}}{\rightarrow} \Gamma(\mathbf{Q},\tau) \stackrel{{\rm FT}}{\rightarrow} \Gamma(\mathbf{r},\tau) \, .
\label{eq:Gamma_FT_scheme}
\end{equation}
Here, it is again necessary to subtract an analytic function that contains the leading behavior for large $(\mathbf{Q},i\Omega)$. 
The aim, in practice, is to obtain a difference function $\Delta \Gamma$ that decays like $\Omega^{-5/2}$ for large $\Omega$, similarly to the difference function of Eq.~(\ref{eq:Gn(k,omega)}) for the single-particle Green's functions. 
To this end, one can make use of the known leading behavior of $R_\mathrm{pp}(\mathbf{Q},i\Omega)$ for large $(\mathbf{Q},i\Omega)$ (cf.~Eq.~(\ref{eq:Rppa_Q_Omega})), together with the definition (\ref{eq:Gamma_Q}) 
of $\Gamma(\mathbf{Q},i\Omega)$ which in dimensionless units reads
\begin{equation}
\Gamma(\mathbf{Q},i\Omega)=\frac{(-4 \pi)}{v-8\pi R_\mathrm{pp}(\mathbf{Q},i\Omega)} \, .
\label{eq:Gamma_Q_Omega}
\end{equation}
From the leading term $E_0$ of $R_\mathrm{pp}^{(a)}$ (cf.~Eq.~(\ref{eq:E0_Q_Omega})), we obtain for the leading behavior
\begin{equation}
\Gamma^{(a,0)}(\mathbf{Q},i\Omega) = \frac{(-4 \sqrt{2} \pi)}{\sqrt{2} v- ( \frac{\gamma_\uparrow \gamma_\downarrow}{2}\mathbf{Q}^2-2 \mu_0 - i \Omega)^{1/2}} \, ,
\label{eq:Gammaa0_Q_Omega}
\end{equation}
which coincides with the expression of the particle-particle propagator in the strong-coupling limit ($v \gg 1$) \cite{Pieri-2000} but now with $\mu_0$ replacing $\mu$. 
This is because it is here essential that the auxiliary chemical potential satisfies the condition $\mu_0<-v^2$, otherwise the denominator of (\ref{eq:Gammaa0_Q_Omega}) may vanish for $\Omega =0$. 
The term $\Delta \mu \, E_1$ of $R_\mathrm{pp}^{(a)}$ (cf.~Eqs.~(\ref{eq:Rppa_r_tau}) and (\ref{eq:E1_Q_Omega})) yields instead the following sub-leading contributions:
\begin{align}
\label{Gammaa1_Q_Omega}
\Gamma^{(a,1)}(\mathbf{Q},i\Omega) &= \frac{4 \sqrt{2}  \pi \Delta \mu}{ ( \frac{\gamma_\uparrow \gamma_\downarrow}{2}\mathbf{Q}^2-2 \mu_0 - i \Omega)^{3/2}} \\
\label{Gammaa2A_Q_Omega}
\Gamma^{(a,2\text{A})}(\mathbf{Q},i\Omega) &= \frac{16 \pi v \Delta \mu}{ (\frac{\gamma_\uparrow \gamma_\downarrow}{2}\mathbf{Q}^2-2 \mu_0 - i \Omega)^{2}} \, .
\end{align}
In addition, while the contribution of the term $[(\Delta \mu)^2-(\Delta h)^2] E_2$ in Eq.~(\ref{eq:Rppa_r_tau}) can be neglected since it is of order $\Omega^{-5/2}$ or higher,
the contribution of $\tilde{R}_\mathrm{pp}^{(sa)}$ (cf.~Eq.~(\ref{eq:Rppsa'_Q_Omega})) that enters the expression (\ref{eq:tildeDRpp_Q_tau}) has instead to be taken into account since it is of order $\Omega^{-2}$:
\begin{equation}
\begin{split}
\Gamma^{(a,2\text{B})}(\mathbf{Q},i\Omega) =\sum_\sigma \frac{(-64 \pi^2 n_{\bar{\sigma}})}{\gamma_\sigma \mathbf{Q}^2- \tilde{\mu}_{0 \sigma}-i \Omega} \\\times \frac{1}{ \frac{\gamma_\uparrow \gamma_\downarrow}{2}\mathbf{Q}^2-2 \tilde{\mu}_0 - i \Omega } \, ,
\end{split}
\label{eq:Gammaa2B_Q_Omega}
\end{equation}
with $\tilde{\mu}_{0,\sigma}= 4 \gamma_\sigma \mu_0$ and $\tilde{\mu}_0 = \gamma_\uparrow \gamma_\downarrow \mu_0$. 
Note that these auxiliary chemical potentials are different from those used in Eq.~(\ref{eq:Rppsa'_Q_Omega}). 
However, the present choice does not modify the leading 
$\Omega^{-2}$ behavior of Eq.~(\ref{eq:Gammaa2B_Q_Omega}) and is useful for computing the Fourier transform analytically. 

At this point, the expressions (\ref{eq:Gammaa0_Q_Omega})-(\ref{eq:Gammaa2B_Q_Omega}) can be Fourier transformed to the $(\mathbf{Q},\tau)$ space, yielding:
\begin{align}
\label{eq:Gammaa0_Q_tau}
\begin{split}
\Gamma^{(a,0)}(\mathbf{Q},\tau) =& \bigg[ \frac{4\sqrt{2 \pi}}{\sqrt{\tau}} c(\tau,v)+16 \pi \, v  \, e^{2 v^2 \tau} \theta(v) \bigg] \\
 & \times e^{-(\frac{\gamma_\uparrow \gamma_\downarrow}{2} \mathbf{Q}^2 - 2\mu_0) \tau} \, \theta(\tau) 
\end{split} \\
\Gamma^{(a,1)}(\mathbf{Q},\tau) =& 8 \sqrt{2 \pi \tau} \, \Delta \mu \, e^{-(\frac{\gamma_\uparrow \gamma_\downarrow}{2} \mathbf{Q}^2 - 2\mu_0) \tau} \, \theta(\tau)  \\
\Gamma^{(a,2\text{A})}(\mathbf{Q},\tau) =&  16 \pi  v \, \tau \, \Delta \mu \,  e^{-(\frac{\gamma_\uparrow \gamma_\downarrow}{2} \mathbf{Q}^2 - 2\mu_0) \tau} \, \theta(\tau)  \\
\label{eq:Gammaa2B_Q_tau}
\begin{split}
\Gamma^{(a,2\text{B})}(\mathbf{Q},\tau) =& -64 \pi^2 \theta(\tau) \sum_\sigma \frac{n_{\bar{\sigma}}/\gamma_\sigma^2}{\frac{\mathbf{Q}^2}{2} - 2 \mu_0}\\  \times \big[ &e^{-\gamma_\uparrow \gamma_\downarrow (\frac{\mathbf{Q}^2}{2} - 2\mu_0) \tau} - e^{-2\gamma_\sigma (\frac{\mathbf{Q}^2}{2} - 2\mu_{0 \sigma}) \tau}  \big] \, ,
\end{split}
\end{align}
where in the last expression we have made use of the relation $2 \tilde{\mu}_0 - \tilde{\mu}_{0 \sigma}=- 2 \gamma_\sigma^2 \mu_0$. 
In Eq.~(\ref{eq:Gammaa0_Q_tau}) the coefficient $c(\tau,v)$ is defined by
\begin{equation}
c(\tau,v)= \frac{2}{\sqrt{\pi}}\int_0^{+\infty}  \! \! \! dx \, e^{-x^2} \frac{x^2}{x^2+2 \tau v^2} \, ,
\label{eq:c_tau_v}
\end{equation}
which equals $1$ when either $v=0$ or $\tau=0$ and can be computed numerically otherwise. 
To obtain the expressions (\ref{eq:Gammaa0_Q_tau})-(\ref{eq:c_tau_v}), we have again relied on the property $\mu_0<-v^2$ of the auxiliary chemical potential.

Finally, the Fourier transforms from $(\mathbf{Q},\tau)$ to $(\mathbf{r}, \tau)$ spaces of $\Gamma^{(a,0)}$, $\Gamma^{(a,1)}$, and $\Gamma^{(a,2\text{A})}$ are readily obtained, in the form:
\begin{align}
\begin{split}
\Gamma^{(a,0)}(\mathbf{r},\tau) =& \bigg[ \frac{2 \, c(\tau, v)}{\pi \sqrt{\tau}} + 4 \sqrt{\frac{2}{\pi}} \, v \, e^{2 v^2 \tau} \theta(v) \bigg] \\
&\times \frac{e^{2 \mu_0 \tau} e^{-\frac{\mathbf{r}^2}{2 \gamma_\uparrow \gamma_\downarrow \tau}}}{(\gamma_\uparrow \gamma_\downarrow \, \tau)^{3/2}} \theta(\tau) \, ,
\end{split} 	\label{eq:Gammaa0_r_tau}  \\
\Gamma^{(a,1)}(\mathbf{r},\tau) =& \frac{4 \, \Delta \mu}{ \pi (\gamma_\uparrow \gamma_\downarrow)^{3/2}}  \frac{e^{2 \mu_0 \tau} e^{-\frac{\mathbf{r}^2}{2 \gamma_\uparrow \gamma_\downarrow \tau}}}{\tau}  \theta(\tau)  	\label{eq:Gammaa1_r_tau}  \, , \\
\Gamma^{(a,2\text{A})}(\mathbf{r},\tau) =& \frac{16 \pi v \, \Delta \mu}{(2 \pi \gamma_\uparrow \gamma_\downarrow)^{3/2}}  \frac{e^{2 \mu_0 \tau} e^{-\frac{\mathbf{r}^2}{2 \gamma_\uparrow \gamma_\downarrow \tau}}}{\sqrt{\tau}} \theta(\tau) \, . 
\label{eq:Gammaa2A_r_tau}
\end{align}
For $\Gamma^{(a,2\text{B})}$, on the other hand, the procedure is somewhat more involved and is based on the identity
\begin{eqnarray}
&&\int \frac{d \mathbf{Q}}{(2 \pi)^3} e^{i \mathbf{Q} \cdot \mathbf{r}} \frac{e^{-\gamma \tau (\mathbf{Q}^2/2-2 \mu_0)}}{\mathbf{Q}^2/2-2\mu_0} = \nonumber \\
&&=\int\!\! \frac{d \mathbf{Q}}{(2 \pi)^3} e^{i \mathbf{Q} \cdot \mathbf{r}} \!\!\int_{\gamma \tau}^{+\infty} \! \! \! \!\!d y \, e^{-y (\frac{\mathbf{Q}^2}{2}-2\mu_0)} = \int_{\gamma \tau}^{+\infty} \! \! \! \!\!d y \, \frac{e^{2\mu_0 y} e^{-\frac{\mathbf{r}^2}{2y}}}{(2\pi y)^{3/2}} \nonumber \\
&&= \frac{e^{-2 \sqrt{|\mu_0|} |\mathbf{r}|}}{4 \pi |\mathbf{r}|} \bigg\{ 1+ \text{Erf}\bigg( \frac{|\mathbf{r}|-2\sqrt{|\mu_0|}\gamma \tau}{\sqrt{2 \gamma \tau}} \bigg)\nonumber\\
&&-e^{4 \sqrt{|\mu_0|} |\mathbf{r}|} \bigg[1 - \text{Erf} \bigg(  \frac{|\mathbf{r}|+2\sqrt{|\mu_0|}\gamma \tau}{\sqrt{2 \gamma \tau}} \bigg) \bigg] \bigg\}
\end{eqnarray}
where $\text{Erf}(x)$ is the error function of real argument $x$ \cite{AS-1972}.
Use of this identity for $\gamma=\gamma_\uparrow \gamma_\downarrow$ or $\gamma=2\gamma_\sigma$ yields:
\begin{widetext} 
\begin{equation}
\label{eq:Gammaa2B_r_tau}
\begin{split}
\Gamma^{(a,2\text{B})}(\mathbf{r},\tau) =& -\frac{16 \pi \theta(\tau)}{|\mathbf{r}|} \sum_\sigma \frac{n_{\bar{\sigma}}}{\gamma_\sigma^2} \bigg\{
e^{-2 \sqrt{|\mu_0|} |\mathbf{r}|} \bigg[ \text{Erf} \bigg(  \frac{|\mathbf{r}|-2\sqrt{|\mu_0|}\gamma_\uparrow \gamma_\downarrow \tau}{\sqrt{2 \gamma_\uparrow \gamma_\downarrow \tau}} \bigg) -  \text{Erf} \bigg(  \frac{|\mathbf{r}|-4\sqrt{|\mu_0|}\gamma_\sigma \tau}{2\sqrt{ \gamma_\sigma \tau}} \bigg) \bigg] \\
&+ e^{2 \sqrt{|\mu_0|} |\mathbf{r}|} \bigg[ \text{Erf} \bigg(  \frac{|\mathbf{r}|+2\sqrt{|\mu_0|}\gamma_\uparrow \gamma_\downarrow \tau}{\sqrt{2 \gamma_\uparrow \gamma_\downarrow \tau}} \bigg) -  \text{Erf} \bigg(  \frac{|\mathbf{r}|+4\sqrt{|\mu_0|}\gamma_\sigma \tau}{2\sqrt{ \gamma_\sigma \tau}} \bigg)
\bigg] \bigg\} \, .
\end{split}
\end{equation}
\end{widetext}

It is now convenient to introduce the difference function
\begin{equation}
\begin{split}
\Delta \Gamma (\mathbf{Q},i\Omega) &= \Gamma ( \mathbf{Q}, i\Omega) - \Gamma^{(a,0)} ( \mathbf{Q},i \Omega) - \Gamma^{(a,1)} ( \mathbf{Q}, i\Omega)\\
& - \Gamma^{(a,2\text{A})} ( \mathbf{Q},i \Omega) - \Gamma^{(a,2\text{B})} ( \mathbf{Q}, i\Omega) \, ,
\end{split}
\label{eq:Delta_Gamma}
\end{equation}
that can be Fourier transformed numerically first over $\Omega$ and then over $\mathbf{Q}$ to obtain $\Delta \Gamma (\mathbf{r},\tau)$. 
The desired $\Gamma (\mathbf{r},\tau)$ then results by adding to this $\Delta \Gamma (\mathbf{r},\tau)$ the analytic expressions (\ref{eq:Gammaa0_r_tau})-(\ref{eq:Gammaa2A_r_tau}) and (\ref{eq:Gammaa2B_r_tau}) obtained above.

\subsection{Transforming from $\Sigma_\sigma(\mathbf{r},\tau)$ to $\Sigma_\sigma(\mathbf{k},i\omega)$}
The last functions to be Fourier transformed are the self-energies $\Sigma_\sigma$. 
Similarly to the previous functions, the Fourier transform is again separated in two steps:
\begin{equation}
\Sigma_\sigma (\mathbf{r}, \tau) \stackrel{{\rm FT}}{\rightarrow} \Sigma_\sigma (\mathbf{k},\tau)\stackrel{{\rm FT}}{\rightarrow} \Sigma_\sigma (\mathbf{k},i\omega) \, .
\label{eq:Sigma_FT_scheme}
\end{equation}
Here, too, it is convenient to identify an analytic function that contains the leading behavior for $\tau\to 0^\pm$ and small ${\bf r}$, so as to subtract it before proceeding to the Fourier transform. 

The starting point is the definition (\ref{eq:Sigma_k}) of the self-energy $\Sigma_\sigma$ in terms of the particle-particle propagator $\Gamma$ and the single-particle Green's function $G_{\bar{\sigma}}$, which in $(\mathbf{r},\tau)$ space and dimensionless units reads
\begin{equation}
\Sigma_\sigma (\mathbf{r}, \tau) = -2 \Gamma(\mathbf{r}, \tau) G_{\bar{\sigma}}(\mathbf{r}, -\tau) \, ,
\label{eq:Sigma_r_tau}
\end{equation}
with the factor $2$ originating from the normalization of $\Gamma$. 
From Sections~\ref{subsec:transforming_G} and \ref{subsec:transforming_Gamma}, both $\Gamma(\mathbf{r}, \tau)$ and $G_{\sigma}(\mathbf{r}, \tau)$ are known to present a singular behavior for $\tau \to 0^+$ (cf.~Eqs.~(\ref{eq:Gammaa0_r_tau}) and (\ref{eq:Ga(r,tau)})). 
For the self-energy (\ref{eq:Sigma_r_tau}), this translates into a singular behavior when both $\tau \to 0^+$ (due to $\Gamma$) and $\tau \to 0^-$ (due to $G_{\bar{\sigma}}$).
Accordingly, we write:
\begin{align}
\Sigma_\sigma (\mathbf{r},\tau) &\underset{\tau \to 0^+}{\simeq} -2 \, G_{\bar{\sigma}}(\mathbf{r}=0, \tau=0^-) \, \Gamma(\mathbf{r},\tau) \\
\Sigma_\sigma (\mathbf{r},\tau) &\underset{\tau \to 0^-}{\simeq} -2 \, G_{\bar{\sigma}}(\mathbf{r}, -\tau) \, \Gamma(\mathbf{r}=0,\tau=0^-) \, ,
\end{align}
where, at the lowest order, we have taken $\mathbf{r}=0$ and $\tau=0^{-}$ in the smooth functions multiplying the singular ones. 
Recalling that the leading singular behaviors of $\Gamma$ and $G_\sigma$ are given by Eqs.~(\ref{eq:Gammaa0_r_tau}) and (\ref{eq:Ga(r,tau)}), we can then use the following functions
\begin{align}
\Sigma_\sigma^{(a,+)} (\mathbf{r},\tau) =& -2 \, n_{\bar{\sigma}} \, \Gamma^{(a,0)}(\mathbf{r},\tau)  \, , \label{eq:Sigmaa+_r_tau} \\
\Sigma_\sigma^{(a,-)} (\mathbf{r},\tau) =& -2 \, \mathcal{C} \, G^{(a)}_{\bar{\sigma}}(\mathbf{r},-\tau)   \, , \label{eq:Sigmaa-_r_tau}
\end{align}
to account for the leading behavior of $\Sigma_\sigma (\mathbf{r},\tau)$ for $\tau\to 0^+$ and  $\tau\to 0^-$, in the order.
Here, $n_{\bar{\sigma}}=G_{\bar{\sigma}}(\mathbf{r}=0, \tau=0^-)$ is the density of the $\bar{\sigma}$-component and $\mathcal{C}=\Gamma(\mathbf{r}=0,\tau=0^-)$ the Tan's contact \cite{Tan-2008}.
These expressions can be readily Fourier transformed to the $(\mathbf{k},i\omega)$ space, yielding
\begin{align}
\Sigma_\sigma^{(a,+)} (\mathbf{k},i\omega) =& -2 \, n_{\bar{\sigma}} \, \Gamma^{(a,0)} (\mathbf{k},i\omega)   \, , \label{eq:Sigmaa+_k_omega} \\
\Sigma_\sigma^{(a,-)}  (\mathbf{k},i\omega)  =& -2 \, \mathcal{C} \, G^{(a)}_{\bar{\sigma}}  (\mathbf{k},-i\omega)   \, , \label{eq:Sigmaa-_k_omega}
\end{align}
where $\Gamma^{(a,0)} (\mathbf{k},i\omega)$ is given by Eq.~(\ref{eq:Gammaa0_Q_Omega}) with $\mathbf{k}$ and  $\omega$ replacing $\mathbf{Q}$ and $\Omega$, respectively, and $G^{(a)}_{\bar{\sigma}}  (\mathbf{k},-i\omega)$ 
is given by Eq.~(\ref{eq:Ga(k,omega)}). 
Out of the two terms (\ref{eq:Sigmaa+_k_omega}) and (\ref{eq:Sigmaa-_k_omega}),  $\Sigma_\sigma^{(a,+)}$ has the strongest singularity.  
For this reason, in our calculation we found it sufficient to define a difference function only in terms of $\Sigma_\sigma^{(a,+)}$, namely,
\begin{equation}
\Delta \Sigma_\sigma (\mathbf{r},\tau) = \Sigma_\sigma(\mathbf{r},\tau)-\Sigma^{(a,+)}_\sigma(\mathbf{r},\tau) \, ,
\label{Delta-Sigma-erre-tau}
\end{equation}
which can be Fourier transformed numerically first over $\mathbf{r}$ and then over $\tau$. 
The desired $\Sigma_\sigma (\mathbf{k},i\omega)$ is finally obtained by adding the expression (\ref{eq:Sigmaa+_k_omega}) of $\Sigma_\sigma^{(a,+)}(\mathbf{k},i\omega)$ to $\Delta \Sigma_\sigma (\mathbf{k},i\omega)$
obtained in this way from Eq.~(\ref{Delta-Sigma-erre-tau}).

\subsection{Numerical grids}
In this Section, we provide details about the grids used in the numerical calculations of the Fourier transforms. 

\noindent
(i) 
For the $\tau$ grid, we use a double logarithmic grid with $600$ points in the interval $(-\tau_\text{max},\tau_\text{max})$ where $\tau_\text{max}=10^8$, which concentrates points around $\tau=0$. 

\noindent
(ii)
For the integration over the frequencies $\omega$ and $\Omega$, we first note that these integrals can be restricted from $0$ to $\infty$ owing to the symmetries 
$G_\sigma(\mathbf{k},-i \omega)=G_\sigma(\mathbf{k},i \omega)^*$ and $\Gamma(\mathbf{Q},-i\Omega)=\Gamma(\mathbf{Q},i\Omega)^*$. 
We use 500 points for both frequencies, where the first $100$ points are taken over an evenly spaced grid with step $\Delta \omega = \Delta \Omega = 2 \pi /\tau_\text{max}$ to correctly describe the low-frequency region, 
while the remaining $400$ points are taken over a logarithmic grid to correctly recover the \mbox{$\sim \omega^{-5/2}$} tail of $G^{(n)}_\sigma(\mathbf{k},i \omega)$ and the $\sim \Omega^{-5/2}$ tail of $\Delta \Gamma(\mathbf{Q},i\Omega)$. The small value of $\Delta \Omega$ for the linear grid and the corresponding large value of $\tau_\text{max}$ are needed to correctly integrate the almost divergence of $\Gamma(\mathbf{Q},i\Omega)$ for $\Omega \to 0$ and $|\mathbf{Q}| \simeq Q_0$, when close to a FFLO QCP (cf.~Eq.~(\ref{eq:Thouless})). 

\noindent
(iii)
For the $\mathbf{k}$ grid, we use $200$ points in total and split the integral in the three intervals $(0, k_{{\rm F}1})$, $(k_{{\rm F}1}, k_{{\rm F}2})$, and $(k_{{\rm F}2}, k_\text{max})$, where $k_{{\rm F}1}$ and $k_{{\rm F}2}$ are, respectively, the smaller and the larger of the two Fermi wave vectors $k_{{\rm F} \sigma}$, and $k_\text{max}$ is a momentum cutoff (for which we typically take $k_\text{max} = 50$ (in units of the original $k_{\rm F}$). 
The points on the grid are arranged such that they are concentrated close to the two Fermi wave vectors $k_{{\rm F} \sigma}$ (albeit avoiding values of $|\mathbf{k}|$ in the region $||\mathbf{k}|-k_{{\rm F} \sigma}|<3\times 10^{-4}$ 
to prevent problems at the discontinuity of the Fermi surfaces), and logarithmically distributed for large $|\mathbf{k}|$. 

\noindent
(iv)
For the integration over $\mathbf{Q}$, we use $300$ points, some of which logarithmically distributed for large $|\mathbf{Q}|$, others concentrated close to $\mathbf{Q}=0$, and the remaining ones concentrated close to 
the (absolute or local) minimum of $\Gamma(\mathbf{Q},i\Omega=0)^{-1}$, when this occurs for $|\mathbf{Q}|\neq 0$.

\noindent
(v)
Finally, for the variable $\mathbf{r}$ we typically use a logarithmic grid with $1000$ points in the interval $(0,r_\text{max})$ where $r_\text{max}=200$ (in units of $k_{\rm F}^{-1}$).

\subsection{Fit of residue curves to extract the crossover polarization $p^*$}
\label{subsec:fitZ}
\begin{figure}[th]    
\includegraphics[angle=0,width=\columnwidth]{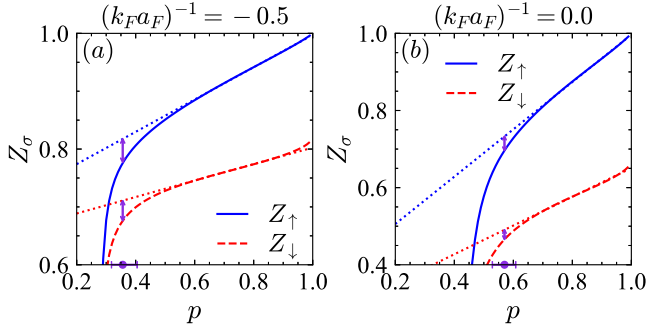}
\caption{ Fit of the linear part of the quasi-particle residue curves $Z_\sigma$ vs the polarization $p$ for the two spin components $\sigma=(\uparrow,\downarrow)$, at zero temperature and for the coupling values $(k_{\rm F} a_{\rm F})^{-1}=(-0.5,0.0)$. The violet double arrows indicate the point where the deviation of the $Z_\sigma(p)$ curves from the linear fit is $5\%$, corresponding to our definition for the crossover polarization $p^*$ (indicated also by the violet circles with error bars on the horizontal axis). }
\label{Figure-10}
\end{figure}

Figure~\ref{Figure-10} shows the fitting procedure used to extract the crossover polarization $p^*$ from the residue $Z_\sigma$ vs polarization $p$ curves (cf.~Fig.~\ref{Figure-5} in the main text). The linear part of the $Z_\sigma(p)$ curves is first fitted with a line (dotted lines in Fig.~\ref{Figure-10}), and then $p^*$ is identified as the polarization at which the deviation of the $Z_\sigma(p)$ curves from the linear fit is $5\%$. Note that the result  for $p^*$ is essentially independent on the spin, so that the same $p^*$ is obtained from the fit of $Z_\uparrow(p)$ and $Z_\downarrow(p)$ curves. The error bars on $p^*$ are defined as the interval of polarizations for which the deviation from the fitted line spans from $2.5\%$ to $10\%$.

\section{A SHORTCOMING OF THE NON-SELF-CONSISTENT $t$-MATRIX FOR THE POLARIZED SYSTEM AT $T=0$}
\label{app:failure_NSC}
In this Appendix, we discuss a shortcoming of the non-self-consistent $t$-matrix approach in the description of a polarized Fermi system at $T=0$.
This consists in not including, for the majority component, the effects of the inter-particle interaction with the the minority component. 
In particular, within this approach the majority component remains non-interacting as long as the chemical potential of the minority component $\mu_\downarrow$ is negative. 
For instance, at unitarity $\mu_\downarrow$ changes sign for polarization $p\simeq 0.38$, implying that in the whole interval $0.38<p<1$ the majority component is treated as completely non-interacting. 
This shortcoming of the non-self-consistent $t$-matrix approach was already pointed out in Ref.~\cite{Schneider-2009}.
Here, we present an analytic proof that accounts for this behavior, by showing that $\Sigma_{0 \uparrow}$ vanishes identically as long as $\mu_\downarrow$ is negative.

Let's consider the expression of self-energy $\Sigma_{0\uparrow}(\mathbf{r},\tau)$ within the non-self-consistent approach. 
Similarly to Eq.~(\ref{eq:Sigma_r_tau}), we write it in the form
\begin{equation}
\Sigma_{0\uparrow} (\mathbf{r}, \tau) = -\, \Gamma_0(\mathbf{r}, \tau) G_{0\downarrow}(\mathbf{r}, -\tau) \, ,
\label{eq:Sigma0_r_tau}
\end{equation}
where $ G_{0\sigma}$ is the free single-particle Green's function for the $\sigma$ component, $\Gamma_0$ is the particle-particle propagator built in terms of $ G_{0\sigma}$ instead of $G_{\sigma}$,
and we have reverted to dimensional expressions.
It is convenient  to separate the two regions $\tau >0$ and $\tau < 0$, by defining
\begin{equation}
\begin{split}
\Sigma_{0\uparrow} (\mathbf{r}, \tau) = \Sigma^{(+)}_{0\uparrow} (\mathbf{r}, \tau) \theta(\tau) +  \Sigma^{(-)}_{0\uparrow} (\mathbf{r}, \tau) \theta(-\tau) 
\label{Bself}
\end{split}
\end{equation}
with
\begin{eqnarray}
&\Sigma^{(+)}_{0\uparrow} (\mathbf{r}, \tau)=- \, \Gamma^{(+)}_0(\mathbf{r}, \tau) G^{(-)}_{0\downarrow}(\mathbf{r}, -\tau)  \\
&\Sigma^{(-)}_{0\uparrow}(\mathbf{r}, \tau) =- \, \Gamma^{(-)}_0(\mathbf{r}, \tau) G^{(+)}_{0\downarrow}(\mathbf{r}, -\tau) \, ,
\end{eqnarray}
and correspondingly
\begin{eqnarray}
&G_{0\downarrow}(\mathbf{r}, \tau)= G^{(+)}_{0\downarrow}(\mathbf{r}, \tau) \theta(\tau) + \, G^{(-)}_{0\downarrow}(\mathbf{r}, \tau) \theta(-\tau) , \\
& \Gamma_0(\mathbf{r}, \tau) = \Gamma_0^{(+)}(\mathbf{r}, \tau) \theta(\tau) +  \Gamma_0^{(-)}(\mathbf{r}, \tau) \theta(-\tau) \, .
\end{eqnarray}

It can be shown that both terms of Eq.~(\ref{Bself}) vanish when $\mu_\downarrow<0$, because in this case $G^{(-)}_{0\downarrow} (\mathbf{r},\tau) =0$ and $\Gamma^{(-)}_0 (\mathbf{r},\tau) =0$. 
For the first term of Eq.~(\ref{Bself}), it is sufficient to recall that 
\begin{equation}
\begin{split}
G_{0 \sigma}(\mathbf{k},\tau)= &- e^{-\xi_{\mathbf{k}\sigma} \tau}\theta(\xi_{\mathbf{k}\sigma}) \theta(\tau) \\
&+ e^{-\xi_{\mathbf{k}\sigma} \tau}\theta(-\xi_{\mathbf{k}\sigma}) \theta(-\tau) 
\end{split}
\end{equation}
where $\xi_{\mathbf{k}\sigma} = \mathbf{k}^2/(2 m) - \mu_\sigma$ (for equal masses). 
When $\mu_\downarrow<0$, one has that $\xi_{\mathbf{k}\downarrow}>0$ for all $\mathbf{k}$, implying also that 
$G^{(-)}_{0\downarrow}(\mathbf{k},-\tau)=e^{\xi_{\mathbf{k}\downarrow} \tau}\theta(-\xi_{\mathbf{k}\downarrow}) = 0$ 
and thus $\Sigma^{(+)}_{0\uparrow} (\mathbf{r}, \tau)=0$. 

The proof that also the second term of Eq.~(\ref{Bself}) vanishes requires some more efforts. 
We consider the spectral representation of $\Gamma_0(\mathbf{Q},i\Omega)$ \cite{Pisani-2004}
\begin{equation}
\Gamma_0(\mathbf{Q},i\Omega) = - \frac{1}{\pi} \int_{-\infty}^{+\infty} \! \! \! d\tilde{\Omega} \, \frac{\text{Im}  \Gamma^\text{R}_0(\mathbf{Q},\tilde{\Omega})}{i \Omega - \tilde{\Omega}} \, ,
\label{eq:Gamma0_spectral_rep}
\end{equation}
where $\Gamma^\text{R}_0(\mathbf{Q},\tilde{\Omega})$ is the retarded particle-particle propagator which is obtained by performing the analytic continuation $i \Omega \to \tilde{\Omega} + i 0^+$ in $\Gamma_0(\mathbf{Q},i\Omega)$,
namely, 
\begin{equation}
\Gamma^\text{R}_0 (\mathbf{Q},\tilde{\Omega}) = - \bigg[\frac{m}{4\pi a_{\rm F}} + R^\text{R}_{\mathrm{pp},0}(\mathbf{Q},\tilde{\Omega}) \bigg]^{-1}
\label{eq:GammaR0_Q}
\end{equation}
with
\begin{equation}
\begin{split}
&R^\text{R}_{\mathrm{pp},0}(\mathbf{Q},\tilde{\Omega}) = \\
& \int \! \! \frac{d \mathbf{k}}{(2\pi)^3} \bigg[ \frac{1-\theta(-\xi_{\mathbf{k}+\mathbf{Q}/2 \, \uparrow})-\theta(-\xi_{\mathbf{k}-\mathbf{Q}/2 \, \downarrow})}{\mathbf{k}^2/m+\mathbf{Q}^2/(4m)-2 \mu -\tilde{\Omega}-i 0^+} - \frac{m}{\mathbf{k}^2} \bigg] \, .
\end{split}
\label{eq:RppR0_Q}
\end{equation}
 By Fourier transforming the expression (\ref{eq:Gamma0_spectral_rep}) to the $\tau$ space, we obtain
\begin{equation}
\begin{split}
&\Gamma_0(\mathbf{Q},\tau)=\int_{-\infty}^{+\infty} \! \! \! \! \! \! d\tilde{\Omega} \, \frac{\text{Im}  \Gamma^\text{R}_0(\mathbf{Q},\tilde{\Omega}) }{\pi} e^{-\tilde{\Omega} \tau}  \\
&\quad \quad \times \big[ \theta(\tau) \theta(\tilde{\Omega})+\theta(-\tau) \theta(-\tilde{\Omega})  \big] \, ,
\end{split}
\end{equation}
from which we isolate the term of interest
\begin{equation}
\Gamma^{(-)}_0(\mathbf{Q},\tau)= \frac{1}{\pi} \int_{-\infty}^{0} \! \! \! \! \! \! d\tilde{\Omega} \,\, \text{Im} \Gamma^\text{R}_0(\mathbf{Q},\tilde{\Omega})  \, e^{-\tilde{\Omega} \tau} ~.
\label{eq:Gamma0-_(Q,tau)}
\end{equation}
This term vanishes if $\text{Im} \Gamma^\text{R}_0(\mathbf{Q},\tilde{\Omega})$ vanishes for $\tilde{\Omega}<0$. 
Correspondingly, to obtain $\text{Im} \Gamma^\text{R}_0(\mathbf{Q},\tilde{\Omega}) \neq0$ we should either have that $\text{Im} R^\text{R}_{\mathrm{pp},0}(\mathbf{Q},\tilde{\Omega}) \neq 0$ or, 
for an infinitesimal $\text{Im} R^\text{R}_{\mathrm{pp},0}(\mathbf{Q},\tilde{\Omega})$, that $\text{Re} [\Gamma^\text{R}_0(\mathbf{Q},\tilde{\Omega})^{-1}] = 0$ for some values of $\tilde{\Omega}$ 
(that depend on $\mathbf{Q}$). 

The latter possibility is excluded for $\tilde{\Omega}<0$ because it would imply that $\text{Re} [\Gamma^\text{R}_0(\mathbf{Q},\tilde{\Omega}=0)^{-1}] < 0$, whereas in the normal phase one knows that
$\text{Re} [\Gamma^\text{R}_0(\mathbf{Q},\tilde{\Omega}=0)^{-1}]$ should remain positive, in accordance with an extended Thouless criterion valid for any given $\mathbf{Q}$. 
The fact that $\text{Re} [\Gamma^\text{R}_0(\mathbf{Q},\tilde{\Omega})^{-1}] = 0$ for some negative values of $\tilde{\Omega}$ implies also that $\text{Re} [\Gamma^\text{R}_0(\mathbf{Q},\tilde{\Omega}=0)^{-1}] < 0$ for $\tilde{\Omega}=0$ 
follows by considering the derivative of the expression (\ref{eq:RppR0_Q}) with respect to $\tilde{\Omega}$ which, when $\mu_\downarrow < 0$, remains positive for all $\tilde{\Omega}$, implying that $\text{Re} [\Gamma^\text{R}_0(\mathbf{Q},\tilde{\Omega})^{-1}]$ is a monotonically decreasing function of $\tilde{\Omega}$.

On the other hand, to have $\text{Im} R^\text{R}_{\mathrm{pp},0}(\mathbf{Q},\tilde{\Omega}) \neq 0$ the denominator of the fraction within brackets in Eq.~(\ref{eq:RppR0_Q}) which contains the infinitesimal term $i 0^+$ should vanish 
for some values of  $\mathbf{k}$ where the numerator remains finite. 
One can readily verify that, for  $\mu_\downarrow < 0$, this numerator is non-vanishing for ${\mathbf k}^2 >(\sqrt{2m\mu_\uparrow}-|\mathbf{Q}|/2)^2$, implying that  $\text{Im} R^\text{R}_{\mathrm{pp},0}(\mathbf{Q},\tilde{\Omega}) \neq 0$ 
for $\tilde{\Omega} > \tilde{\Omega}_{\rm th} (\mathbf{Q})$ with
 \begin{eqnarray}
\tilde{\Omega}_{\rm th} (\mathbf{Q})&=& \mathbf{Q}^2/(4m)-2\mu +(\sqrt{2m\mu_\uparrow}-|\mathbf{Q}|/2)^2/m \nonumber \\ 
&=& |\mathbf{Q}|^2/(2m)- \sqrt{2 \mu_\uparrow/m} |\mathbf{Q}| + \mu_\uparrow- \mu_\downarrow \, ,
\label{eq:Omega'_th}
\end{eqnarray}
which is always positive when $\mu_\downarrow<0$. 
We then conclude that, when $\mu_\downarrow<0$, $\text{Im} R^\text{R}_{\mathrm{pp},0}(\mathbf{Q},\tilde{\Omega})$ and thus $\text{Im} \Gamma^\text{R}_{0}(\mathbf{Q},\tilde{\Omega})$ vanish for $\tilde{\Omega}<0$.
 
From Eq.~(\ref{eq:Gamma0-_(Q,tau)}) this result, in turn, implies that $\Gamma^{(-)}_0(\mathbf{Q},\tau)=0$, and thus that also $\Gamma^{(-)}_0(\mathbf{r},\tau)=0$ for its Fourier transform. 
Accordingly, also the second term in Eq.~(\ref{Bself}) vanishes, thereby proving our statement that $\Sigma_{0 \uparrow}=0$ identically.
As a corollary, in this way we have also proved that the Tan's contact $\mathcal{C}_0=m^2 \Gamma_0(\mathbf{r}=0,\tau \to 0^-)$ remains zero as long as $\mu_\downarrow<0$.



\begin{thebibliography}{99}
\bibitem{footnote-Z-from-imaginary-part} In Eq.~(\ref{eq:residue}) the residue $Z_{\sigma}$ is obtained from the imaginary part of the self-energ $\Sigma_{\sigma} (k_{{\rm F} \sigma},i\omega)$,
                                                                 because the (partial) derivative is here taken with the frequency $\omega$ varying along the imaginary axis.
\bibitem{footnote_anomalous_dim} In the present fully self-consistent $t$-matrix approach, the exponent 2 of the $(|\mathbf{Q}|-Q_0)^2$ term could in  principle be modified to $2-\eta$ due to a non-analytic behavior introduced 
                                                         by self-consistency. However, our numerical calculations at the QCP show that the anomalous dimension parameter $\eta$ is compatible with zero.
\end{thebibliography}
\end{document}